\documentclass[12pt]{article}
\pdfoutput=1
\usepackage{comment}
\usepackage{amsmath}
\usepackage{amssymb}
\usepackage{graphicx,color}

\usepackage{here}

\usepackage{subcaption}
\usepackage{url}

\usepackage[sort&compress, numbers, merge]{natbib}

\usepackage{braket}
\usepackage{physics}
\usepackage[compat=1.1.0]{tikz-feynhand}
\usepackage{slashed}
\usepackage{mathtools}
\usepackage{bbold}
\usepackage{slashed}
\usepackage{colortbl,array,xcolor}

\usepackage{stackengine}

\numberwithin{equation}{section}

\usepackage[colorlinks=true, linkcolor=blue, citecolor=blue,
urlcolor=black]{hyperref} 

\usepackage{cancel}
\usepackage{diagbox}

\usepackage{tikz}
\usetikzlibrary{positioning,intersections,calc,arrows.meta,matrix,shapes,fit,tikzmark}
\usepackage{tikz-feynhand}

\def\arrowscale{2}

\def\biarrow{\draw[>->,>={Classical TikZ Rightarrow[scale=\arrowscale]}]}
\def\fundarrow{\draw[->,>={Classical TikZ Rightarrow[scale=\arrowscale]}]}
\def\antiarrow{\draw[>-,>={Classical TikZ Rightarrow[scale=\arrowscale]}]}
\def\inwardarrow{\draw[{>[red]}-<,>={Classical TikZ Rightarrow[scale=\arrowscale]}]}
\def\outwardarrow{\draw[{<[red]}->,>={Classical TikZ Rightarrow[scale=\arrowscale]}]}
\def\antiarrowcolored{\draw[>-,>={Classical TikZ Rightarrow[scale=\arrowscale,{red}]}]}
\def\SU{\mathrm{SU}}
\def\SO{\mathrm{SO}}
\def\xE{x_\mathrm{E}}
\def\gE{g^\mathrm{E}}
\def\AE{A_\mathrm{E}}
\def\partialE{\partial_\mathrm{E}}
\def\sigmaE{\sigma_{\mathrm{E}}}
\def\sigmaM{\sigma_{\mathrm{M}}}
\def\barsigmaE{\bar{\sigma}_\mathrm{E}}
\def\barsigmaM{\bar{\sigma}_\mathrm{M}}
\def\barsigma{\bar{\sigma}}
\def\bartau{\bar{\tau}}
\def\LE{\mathcal{L}_{\mathrm{E}}}
\def\boldtau{\boldsymbol{\tau}}
\def\boldsigma{\boldsymbol{\sigma}}
\def\simgt{\mathrel{\lower2.5pt\vbox{\lineskip=0pt\baselineskip=0pt
           \hbox{$>$}\hbox{$\sim$}}}}
\def\simlt{\mathrel{\lower2.5pt\vbox{\lineskip=0pt\baselineskip=0pt
           \hbox{$<$}\hbox{$\sim$}}}}
\def\simprop{\mathrel{\lower3.0pt\vbox{\lineskip=1.0pt\baselineskip=0pt
             \hbox{$\propto$}\hbox{$\sim$}}}}
\newcolumntype{R}{>{\columncolor{red!20}}c}
\newcolumntype{B}{>{\columncolor{cyan!20}}c}
\newcolumntype{Y}{>{\columncolor{yellow!20}}c}
\newcolumntype{G}{>{\columncolor{green!20}}c}
\newcolumntype{O}{>{\columncolor{orange!20}}c}

\begin{document}

\begin{titlepage}

\begin{flushright}
IPMU24-0017	
\end{flushright}

\vskip 1.1cm

\begin{center}

{\Large \bf 
Small Instanton Effects on Composite Axion Mass
\\
}

\vskip 1.2cm
Takafumi Aoki$^{a}$, 
Masahiro Ibe$^{a,b}$,
Satoshi Shirai$^{b}$, and
Keiichi Watanabe$^{a}$ 
\vskip 0.5cm

{\it

$^a$ {ICRR, The University of Tokyo, Kashiwa, Chiba 277-8582, Japan}

$^b$ {Kavli Institute for the Physics and Mathematics of the Universe
 (WPI), \\The University of Tokyo Institutes for Advanced Study, \\ The
 University of Tokyo, Kashiwa 277-8583, Japan}

}

\vskip 1.0cm

\abstract{
This paper investigates the impact of small instanton effects on the axion mass in composite axion models.
In particular, we focus on the Composite Accidental Axion (CAA) models, which are designed to address the axion quality problem, and where the Peccei-Quinn (PQ) symmetry emerges accidentally.
In the CAA models, the QCD gauge symmetry is embedded in a larger gauge group at high energy. 
These models contain small instantons not included in low-energy QCD, which could enhance the axion mass significantly.
However, in the CAA models, our analysis reveals that these effects on the axion mass are non-vanishing but are negligible compared to the QCD effects.
The suppression of the small instanton effects originates from the global chiral U(1) symmetries which are not broken spontaneously and play a crucial role in eliminating $\theta$-terms in the hidden sectors through anomalies.
We find these U(1) symmetries restrict the impact of small instantons in hidden sectors on the axion mass.
Our study provides crucial insights into the dynamics within the CAA models and suggests broader implications for understanding small instanton effects in other composite axion models.}

\end{center}
\end{titlepage}

\tableofcontents

%-----------------------------------------------------
\section{Introduction} 
%-----------------------------------------------------

The Peccei-Quinn (PQ) mechanism is the most prominent solution to the strong CP problem~\cite{Peccei:1977hh,Peccei:1977ur,Weinberg:1977ma,Wilczek:1977pj}. 
In this mechanism, a global U(1) symmetry called PQ symmetry, plays a crucial role. 
This symmetry is an exact symmetry, except for the axial anomaly of QCD. 
With the spontaneous symmetry breaking (SSB) of PQ symmetry, the effective $\theta$--angle of QCD is cancelled by the vacuum expectation value (VEV) of the axion, so that the strong CP problem is solved.

However, the assumption of such a convenient global symmetry seems to be weakly grounded.
By its definition, the PQ symmetry is an inexact symmetry due to the anomaly. 
It is also argued that all global symmetries are broken by quantum gravity effects~\cite{Hawking:1987mz,Lavrelashvili:1987jg,Giddings:1988cx,Coleman:1988tj,Gilbert:1989nq,Banks:2010zn}. 
These arguments suggest that the presence of (higher-dimensional) operators that violate the PQ symmetry cannot generally be ruled out.
When such operators are present, the resultant effective $\theta$--angle easily exceeds the experimental constraints, and spoils the PQ mechanism \cite{Kamionkowski:1992mf,Banks:2010zn,Holman:1992us}. 
This issue is known as the axion quality problem.

The axion quality problem has motivated various extensions of axion models, so that a PQ symmetry emerges as an almost exact symmetry. 
For instance, the accidental PQ symmetry can be achieved by certain (discrete) gauge symmetries, as detailed in Refs.\,\cite{Lazarides:1982tw, Barr:1992qq,Holman:1992us,Dias:2002hz,Carpenter:2009zs,Lazarides:1985bj,Choi:2009jt,Harigaya:2013vja,Gherghetta:2016fhp,Fukuda:2017ylt,Fukuda:2018oco,Ibe:2018hir}. 
Alternatively, a high-quality PQ symmetry may arise from an extra-dimensional setup \cite{Cheng:2001ys,Izawa:2002qk,Hill:2002kq,Fukunaga:2003sz,Izawa:2004bi,Choi:2003wr,Grzadkowski:2007xm}. 
In this paper, we focus on composite axion models 
with high-quality accidental PQ symmetry.

The composite axion model has been proposed to solve another problem of the invisible axion model, i.e., the fine-tuning problem of the axion decay constant.
In the invisible axion model, the astrophysical constraints and dark matter abundance imply
the axion decay constant $f_a$ in the range of  
$f_a = \order{10^{8\mbox{--}12}}$\,GeV
(see e.g.
Ref.\,\cite{Workman:2022ynf}),
which requires fine-tuning.
The composite axion model resolves this issue by achieving the intermediate scale through hidden strongly-coupled gauge dynamics via dimensional transmutation~\cite{Kim:1984pt,Choi:1985cb}.
This hidden gauge interaction, referred to as ``axicolor," confines fermions charged under it, resulting in the axion as a composite state.

In addition to solving the fine-tuning problem, composite axion models are advantageous in achieving high-quality PQ symmetry as an accidental global symmetry of the axicolor fermions.
Several composite axion models have been proposed to address the axion quality problem~\cite{Randall:1992ut,Dobrescu:1996jp,Redi:2016esr,Lillard:2017cwx,Lillard:2018fdt,Gavela:2018paw,Vecchi:2021shj,Contino:2021ayn}.

In many axion models with high-quality PQ symmetry, $\SU(3)_\mathrm{QCD}$ is embedded in a larger gauge group.
In such cases, the axion can couple to the larger gauge group, in addition to $\SU(3)_\mathrm{QCD}$. 
Such dynamics can affect the axion potential, possibly spoiling the PQ mechanism due to the $\theta$-terms in hidden sectors
or modifying the prediction of the axion mass.
Therefore, in extended axion models, effects on the axion potential from these new interactions must be examined.

Among the possible effects from these new interactions, 
instanton configurations not present in QCD
can significantly impact the
axion potential.
These configurations are often called small instantons.
In fact, Ref.\,\cite{Agrawal:2017ksf}
pointed out that
small instanton effects from new gauge interactions can enhance the axion mass
in some models.
The small instanton effects are currently subjects of intense interest and debate in the field \cite{Fuentes-Martin:2019bue,Csaki:2019vte,Gherghetta:2020keg,Kitano:2021fdl,Csaki:2023ziz,Bedi:2024wqg}.

In this paper, we discuss small instanton effects on the axion mass in composite axion models with high-quality PQ symmetry.
In particular, we mainly focus on Composite Accidental Axion (CAA) models, proposed by Redi and Sato \cite{Redi:2016esr}.
One advantageous feature of the CAA models is that the alignment of the effective QCD $\theta$-angle to $0$ is protected by symmetries. 
Specifically, symmetries which are anomalous but not broken spontaneously
eliminate $\theta$-angles in hidden sectors (see Sec.\,\ref{sec:CAA}).
However, it is still unclear whether small instanton effects impact the axion mass, and if they do, to what extent they have an impact.

If the mass of the axion is enhanced, it has substantial theoretical and phenomenological implications.
Theoretically, an increase in axion mass can mitigate the axion quality problem.
Phenomenologically, a higher axion mass significantly impacts accelerator experiments and astrophysical constraints.
For instance, a sufficiently large axion mass can relax experimental and observational constraints, allowing for a smaller decay constant $f_a$,  which further alleviates the axion quality problem~\cite{Fukuda:2015ana}.
Therefore, a detailed study of the axion mass is highly important.

As we will see, small instantons in the CAA models do contribute to the axion mass, in addition to the QCD contributions.
We will also find, however, that those contributions are negligibly small compared with the QCD contributions.
To derive these conclusions, accidental chiral U$(1)$ symmetries
other than the PQ symmetry, which are responsible for eliminating $\theta$-terms in hidden sectors,
play a crucial role.
As discussed in Sec.\,\ref{sec:SIE vs fermions},
the suppression of small instanton effects on the axion mass stems from the fact that the PQ symmetry is not unique but can be redefined as a linear combination with the $\mathrm{U}(1)$ symmetries.
Our conclusion is not limited to the CAA models but is applicable to a wide range of axion models with high-quality PQ symmetry.
It highlights the challenge of enhancing the axion mass without spoiling the high-quality PQ symmetry.

The organization of this paper is as follows.
In Sec.\,\ref{sec:without fermion}, we briefly review the small instanton effects on the axion potential, in a simple model without fermions. 
In Sec.\,\ref{sec:CAA}, we review the composite axion model in Refs.\,\cite{Kim:1984pt,Choi:1985cb} and the simplest CAA model.
In Sec.\,\ref{sec:SIE vs fermions}, we discuss small instanton effects in a perturbative toy model mimicking the CAA model.
In Sec.\,\ref{sec:SIE in CAA}, we discuss the small instanton effects in the CAA model.
In Sec.\,\ref{sec:SIE in other models}, we also discuss other composite axion models.
The final section is devoted to our conclusions.

%-----------------------------------------------------
\section{Small Instanton Effects on Axion Mass without Fermion}
\label{sec:without fermion}
%-----------------------------------------------------
In the CAA models, the accidental PQ symmetry is spontaneously broken due to the axicolor dynamics. This dynamics also breaks a product gauge group, resulting in $\SU(3)_\mathrm{QCD}$ emerging as an unbroken diagonal subgroup.  In this section, we review a model
where small instantons not included in low-energy QCD enhance the axion mass significantly~\cite{Agrawal:2017ksf,Csaki:2019vte}.

%-----------------------------------------------------
\subsection{Constrained Instanton}
\label{sec:constrained instanton}
%-----------------------------------------------------
We consider a model where the small instantons  which are not included in the low-energy QCD
reside in a broken part of the product gauge group. 
Instanton configurations in the broken phase are called constrained instanton~\cite{Affleck:1980mp}.
The essential points of the constrained instantons can be understood by considering a spontaneously broken $\SU(2)$ gauge theory as an example. 

Following Ref.\,\cite{Espinosa:1989qn},
we assume $\SU(2)$ gauge symmetry is broken by the VEV of an $\SU(2)$ doublet scalar field $H$. 
The Euclidean Lagrangian of this system is given by,%
\footnote{We take the Euclidean metric to be $g_{\mu\nu} = (+,+,+,+)$ }
\begin{align}
    \mathcal{L}_{\mathrm{E}} = \dfrac{1}{2 g^2}\mathrm{Tr}(F_{\mu\nu}F_{\mu\nu}) + (D_{\mu}H)^{\dagger}(D_{\mu}H) + \dfrac{\lambda}{4}(H^{\dagger}H - v^2)^2\ ,
    \label{eq:Lagrangian for SU(2) gauge theory with Higgs}
\end{align}
where the field strength and the covariant derivatives are defined by
\begin{align*}
    &
    F_{\mu\nu} = \partial_{\mu}A_{\nu} - \partial_{\nu}A_{\mu} - i\left[A_{\mu},A_{\nu}\right]\ ,
    \\&
    D_{\mu} = \partial_{\mu} - iA_{\mu}\,.
\end{align*}
See Appendix~\ref{app:notations} for notations of Euclidean space.
Here, $g$ is the $\SU(2)$ gauge coupling, $\lambda>0$ is the quartic coupling, 
$v>0$ is a mass dimensionful 
parameter,
and $A_\mu = A_\mu^a \tau^a/2$ ($a=1,2,3$) with $\tau^a$ being the Pauli matrices.
At the vacuum, $H$ obtains a VEV,
\begin{align}
    \langle{H}\rangle = \mqty(0\\v)\ ,
\end{align}
which breaks $\SU(2)$ completely.
The physical scalar boson mass is given by $m_H = \sqrt{\lambda} v$ and the gauge boson mass is given by $m_A = gv/\sqrt{2}$.

For $v = 0$, i.e., in the unbroken phase, 
instanton configurations are local minima of the Euclidean action, which are labeled by winding number $w$.
An instanton configuration with $w=1$ is given by,
\begin{align}
    A_\mu = \dfrac{2\rho^2 x_\nu}{x^2(x^2+\rho^2)} \bartau_{\mu\nu}\ ,\ \ \ \ \ H= 0  \ ,
    \label{eq:instanton w/o Higgs}
\end{align}
where $\rho$ is the instanton size and $\bartau_{\mu\nu}$ is defined in Eq.\,\eqref{eq:generators}.
In the semi-classical approximation,
an instanton configuration contributes to the path integral, which is suppressed by a factor $e^{-S_\mathrm{E}}$ with $S_\mathrm{E} = {8\pi^2}/{g^2}$, which is independent of the size $\rho$.

In the broken phase, where $v \neq 0$, no instanton solutions exist in the strict sense as local minima of the Euclidean action.
As discussed in Ref.\,\cite{Affleck:1980mp}, an instanton-like configuration, whose size is much smaller than the inverse of the symmetry breaking scale $v^{-1}$, contributes to the semi-classical approximation of the path integral, similar to the instantons in the unbroken phase.
An instanton-like configuration with a fixed size is called constrained instanton.

The constrained instanton solution looks like instanton for $x \ll m_{A,H}^{-1}$ 
while decaying 
exponentially at 
$x \gg m_{A,H}^{-1}$. At $x \ll m_{A,H}^{-1}$, the solution behaves as
\begin{align}
\label{eq:constrained instanton profile of A}
    A^\mathrm{inst}_{\mu} &= \dfrac{2\rho^2}{x^2(x^2+\rho^2)} \bartau_{\mu\nu} x_\nu\ + \order{(\rho m_{A,H})^2}\ ,\\
\label{eq:constrained instanton profile of H}
    H^\mathrm{inst} &= \left(\dfrac{x^2}{x^2+\rho^2} \right)^{1/2}
    \begin{pmatrix}
        0\\v
    \end{pmatrix} + \order{(\rho m_{A,H})^2}\ .
\end{align}
Note that $H$ has a nontrivial profile and $\SU(2)$ symmetry is restored at the origin.
Similarly, we also obtain constrained anti-instanton solution by replacing $\bar{\tau}_{\mu\nu}$ with 
$\tau_{\mu\nu}$.

By substituting 
the constrained (anti-)instanton
into the Euclidean action, we obtain,
\begin{align}
    S_\mathrm{E}(\rho) = \frac{8\pi^2}{g^2}+ 
    \frac{4\pi^2\rho^2m_A^2}{g^2} + \frac{1}{g^2}\order{\rho^4 m_{A,H}^4}\ ,
    \label{eq:action w/ constrained instanton}
\end{align}
where the second term comes from the
kinetic term of the scalar field, $|D_\mu H|^2$.
Thus, for a sufficiently small constrained instanton i.e., $\rho m_{A,H} \ll 1$, it can play an important role in the semi-classical approximation
as the (anti-)instanton configuration in the unbroken phase. 

In this section, we have focused on an $\SU(2)$ gauge theory as a simple example. The insights gained about constrained instantons can be applicable to models that include larger gauge groups.

%-----------------------------------------------------
\subsection{Axion Mass Enhancement in Model without Fermion}
\label{sec:without fermion SIE}
%-----------------------------------------------------
To see how the small instantons 
affect the axion potential, let us
consider a model
where $\SU(3)_1\times \SU(3)_2$ 
is broken down to the diagonal subgroup $\SU(3)_\mathrm{QCD}$~\cite{Agrawal:2017ksf}. 
We introduce two axions $a_i$ ($i=1,2$) 
which couple to the $F\tilde{F}$ terms, i.e.
\begin{align}
    \mathcal{L} 
    =  \sum_{i=1}^2 
    \left[-\frac{1}{2g_i^2} 
    \Tr(F_i^{\mu\nu}F_{i\mu\nu})
    + \frac{1}{16\pi^2}  
    \qty(\frac{a_i}{f_i}+\theta_i) \Tr(F_i^{\mu\nu}\tilde{F}_{i\mu\nu})\right]\ .
\end{align}
Here, $i = 1,2$ denotes each gauge sector with the gauge coupling $g_i$ and the
vacuum angle $\theta_i$. 
The domains of the axions 
are given by $a_i/f_i = [0,2\pi)$, with $f_i$ denoting the decay constants of the axions.
In this model, the PQ symmetries are realized as shifts of 
the axions, which are anomalous 
with respect to each $\SU(3)_i$.
In the presence 
of the two axions, both $\theta_1$ and $\theta_2$ can be set to zero by the shifts of the axions.

Let us assume that $\SU(3)_1\times \SU(3)_2$ is broken down to $\SU(3)_{\mathrm{QCD}}$ by the VEV of a bi-fundamental 
($\mathbf{3}$, $\bar{\mathbf{3}}$) scalar field, $\phi_{c}{}^{c'}$ ($c,c'=1,2,3$), i.e.,
\begin{align}
    \langle \phi_c{}^{c'} \rangle = v \delta_{c}{}^{c'}\ .
\end{align}
We also assume that $\SU(3)_1\times \SU(3)_2$ is weakly coupled at the scale around $v$.
Below the breaking scale, 
the effective QCD 
coupling is given by,
\begin{align}
\label{eq:gauge couplings}
    \frac{1}{g_\mathrm{QCD}^2(v)} = \frac{1}{g_1^2(v)}
    +\frac{1}{g_2^2(v)}\ ,
\end{align}
where we take the renormalization scale $\mu = v$.
The axions couple to QCD through 
\begin{align}
    \mathcal{L} 
    = -\frac{1}{2g_\mathrm{QCD}^2} 
    \Tr(G^{\mu\nu}G_{\mu\nu})
    + \frac{1}{16\pi^2}  
    \qty(\sum_{i}\frac{a_i}{f_i} ) \Tr(G^{\mu\nu}\tilde{G}_{\mu\nu})\ ,
\end{align}
where $G_{\mu\nu}$ denotes the field strength of QCD.
Through this coupling, QCD contributes to the axion potential, which is roughly represented by the following expression:
\begin{align}
    V_\mathrm{QCD}(a_1,a_2) \sim \Lambda_\mathrm{QCD}^4 \cos\left(
    \sum_i\frac{a_i}{f_i}
    \right) \ .
    \label{eq:QCD axion potential} 
\end{align}
Here, $\Lambda_\mathrm{QCD}$ is the dynamical scale of QCD.%
\footnote{
In the presence of light quarks,
$V_\mathrm{QCD}\simeq m_\pi^2 f_\pi^2 \cos(a/F_a)$, where $m_\pi$ and $f_\pi$ are the mass and the decay constant of the pion.
Numerically $m_\pi^2 f_\pi^2 \sim\Lambda_\mathrm{QCD}^4$. 
Therefore, for simplicity in the following discussion, we represent the low-energy QCD contributions to the axion potential using
$\Lambda_\mathrm{QCD}^4$.
}
The impact of quarks in the Standard Model (SM) on the axion potential will be discussed at the end of this section.

We denote the winding numbers of $\SU(3)_1$ and $\SU(3)_2$ sectors by $k_1$ and $k_2$.
Following Ref.\,\cite{Csaki:1998vv},
we label them by
$(k_1,k_2)$.
In the semi-classical approximation,
the constrained instanton with a size $\rho$
contributes to the path integral
which is suppressed by $e^{-S_\mathrm{E}}$ 
with the action,
\begin{align}
    S_{\mathrm{E}i}(\rho) = \frac{8\pi^2}{g_i^2(\rho^{-1})}+ 
    \order{4\pi^2\rho^2 v^2} \ ,
\end{align}
where we consider either $(1,0)$
or $(0,1)$ instantons.
Here, we have taken the renormalization scale of the gauge couplings to be the inverse of the small instanton size, $\mu= \rho^{-1}$~\cite{tHooft:1976snw}.

In this simple model, the small constrained instantons generate additional axion potentials,
which are dominated by the contributions from those with $\rho\sim v^{-1}$.
As a result, the total axion potential amounts to 
\begin{align}
    V(a_1,a_2) 
    \sim
    \Lambda_\mathrm{QCD}^4 \cos\left({\sum_{i}\frac{a_i}{f_i}}\right)
    +
    v^4\sum_{i} e^{-\frac{8\pi^2}{g_i^2(v)}}
    \cos\left(\frac{a_i}{f_i}\right)\ .
\end{align}
Since the dynamical scale of QCD is related to the symmetry breaking scale via
\begin{align}
    \Lambda_\mathrm{QCD} = v e^{-\frac{1}{b_{\mathrm{QCD}}}\frac{8\pi^2}{g_{\mathrm{QCD}}^2(v) }}\ ,
\end{align}
the axion potential can be rewritten as
\begin{align}
    V(a_1,a_2)
    \sim
    v^4
    e^{-\frac{4}{b_\mathrm{QCD}}\frac{8\pi^2}{g_\mathrm{QCD}^2(v)}}\cos\left({\sum_{i}\frac{a_i}{f_i}}\right)
    +
    v^4
    \sum_{i}
    e^{-\frac{8\pi^2}{g_i^2(v)}} \cos\left(\frac{a_i}{f_i}\right).
    \label{eq:enhancement without fermion}
\end{align}
Here, $b_{\mathrm{QCD}}$ is the coefficient of the one-loop $\beta$-function of the QCD gauge coupling.

Notably, the additional contributions from the small instantons lead to an increase in the axion mass.
For example, when $g_1= g_2$, we find 
\begin{align}
    g_1^2(v) = g_2^2(v) = 2 g_{\mathrm{QCD}}^2(v) \ ,
\end{align}
and hence, the additional contributions become comparable to the QCD contribution.
If we extend the gauge group 
from $\SU(3)_1\times \SU(3)_2$
to 
$[\SU(3)]^{n_s}$ ($n_s \gg 1$), 
where 
$ \SU(3)_{\mathrm{QCD}}$ appears as the 
diagonal subgroup of $[\SU(3)]^{n_s}$,
the gauge coupling constant at each sector can be as large as $g_i^2(v) \sim n_s g^2_{\mathrm{QCD}}(v)$.
In such cases, further enhancement of the axion mass is possible~\cite{Agrawal:2017ksf}.

In this discussion, we have neglected the effects of SM quarks for simplicity.
Around (constrained) instantons of both QCD and the broken part, SM quarks exhibit zero modes.
Consequently, instanton effects are suppressed by Yukawa coupling constants of SM quarks.
Despite this suppression, however, 
small instanton effects 
can still lead to the enhancement of the axion mass, especially in the cases where
$n_s \gg 1$~\cite{Agrawal:2017ksf,Csaki:2019vte}.

%-----------------------------------------------------
\section{Model of Composite Accidental Axion}
\label{sec:CAA}
%-----------------------------------------------------
In this section, we review the composite axion model in Refs.\,\cite{Kim:1984pt,Choi:1985cb} and the simplest CAA model in Ref.\,\cite{Redi:2016esr}. A key difference of this model from the example in Sec.\,\ref{sec:without fermion SIE} is the presence of new fermions.
As we will see in Secs.\,\ref{sec:SIE vs fermions} and \ref{sec:SIE in CAA},
chiral symmetries of the new fermions
reduce the small constrained instanton effects on the axion mass.

%-----------------------------------------------------
\subsection{Composite Axion Model}
\label{sec:CAM}
%-----------------------------------------------------
\begin{figure}[t]
\centering
\begin{tikzpicture}[scale=2.2]
    \tikzset{t/.style={draw,circle,minimum width=34,opacity=0
    }}
    \tikzset{c/.style={draw,circle,minimum width=32,
    }}
    \node [c](0) at (-1,0){\tiny$\SU(3)$};
    \node [c](1) at (-.707,.707){\tiny$\SU(N)$};
    \node [c](-1) at (-.707,-.707){\tiny$\SU(N)$};
    \node [c](2) at (0,1){\tiny$\SU(4)$};
    \node [c](-2) at (0,-1){\tiny$\SU(4)$};
    \node [c](3) at (.707,.707){\tiny$\SU(N)$};
% transparent nodes
    \node [t](-3) at (.707,-.707){\tiny$\SU(N)$};
    \node [t](4) at (1,0){\tiny$\SU(4)$};
    \node [draw,rectangle](sm) at (-2,0){\tiny SM quarks\,};    
% a bit larger transparent nodes (purpose: keeping arrows from touching nodes)
    \node [](1L) at (-.707-.5,.707){};
    \node [](-1L) at (-.707-.5,-.707){};
    
    \node [t](0) at (0){};
    \node [t](1) at (1){};
    \node [t](-1) at (-1){};
    \node [t](2) at (2){};
    \node [t](-2) at (-2){};
    \node [t](3) at (3){};
    \node [t](-3) at (-3){};  
    \node [t](4) at (1,0){};
    
% end points of dotted curve
    \node [opacity=0](4') at (4){};
    \node [opacity=0](-3') at (-3){};
% dotted curve
    \draw[loosely dashed, thick, domain=.707:1]plot(\x,{-sqrt(1-(\x)^2)});
    \biarrow (0)--(1);
    \biarrow (1)--(2);
    \biarrow (2)--(3);
    \biarrow (3)--(4);
    \biarrow (-3)--(-2);
    \biarrow (-2)--(-1);
    \biarrow (-1)--(0);

    \fundarrow (1L)--(1);
    \antiarrow (-1)--(-1L);

    \draw[dashed](sm)--(0);
\end{tikzpicture}
    \caption{
    The moose diagram of the $\SU(3) \times [\SU(N)]^{n_s} \times [\SU(4)]^{n_s-1}$ composite accidental axion model.
    The arrow \protect\tikz \protect\draw[->,>={Classical TikZ Rightarrow[scale=2.2]}] (0,0)--(0.5,0);
    denotes the fundamental representation  fermion and the arrow \protect\tikz \protect\draw[>-,>={Classical TikZ Rightarrow[scale=2.2]}] (0,0)--(0.5,0);
    the antifundamental representation
    of the left-handed .
    We show tables of the model 
    for  $n_s =2$ and $3$ in Tabs.\,\ref{tab:model n2}
    and \ref{tab:model n3}, respectively.
 } 
 \label{fig:moose CAA}
\end{figure}

In the original composite axion model proposed in Refs.\,\cite{Kim:1984pt,Choi:1985cb}, 
a new $\SU(N)$ confining
gauge interaction called axicolor is introduced.
Hereafter, the confining scale is assumed to be much higher than the electroweak scale.
This model has 
left-handed Weyl fermions charged under $\SU(N)$ and $\SU(3)_\mathrm{QCD}$ as,
\begin{align}
\label{eq:original composite}
    (\mathbf{N},\mathbf{3})_{1} \oplus
    (\overline{\mathbf{N}},\overline{\mathbf{3}})_{1} \oplus
    (\mathbf{N},\mathbf{1})_{-3} \oplus
    (\overline{\mathbf{N}},\mathbf{1})_{-3} \ .
\end{align}
Here, the subscripts indicate the charges under the U$(1)_\mathrm{PQ}$ symmetry, which is imposed by hand. In the limit of vanishing QCD coupling, this model exhibits a chiral flavor symmetry of $\SU(4)_L\times\SU(4)_R\times$U$(1)_V$.%
\footnote{The flavor symmetry of $\mathbf{N}$ fermions is denoted by $\SU(4)_L$ and that of $\overline{\mathbf{N}}$ by $\SU(4)_R$.}
Within this framework,
$\SU(3)_\mathrm{QCD}$ is embedded in 
the vector-like subgroup, i.e., $\SU(3)_\mathrm{QCD} \subset \SU(4)_V \subset \SU(4)_L \times \SU(4)_R$. The U$(1)_\mathrm{PQ}$ symmetry is identified as an axial subgroup of $\SU(4)_L\times \SU(4)_R$, and is therefore anomaly-free with respect to $\SU(N)$.
Note that U$(1)_\mathrm{PQ}$ is, on the other hand, anomalous with respect to $\SU(3)_\mathrm{QCD}$.

At the dynamical scale of $\SU(N)$, $\Lambda$, 
the confinement occurs 
and the (approximate) chiral flavor symmetry is spontaneously broken, i.e.,
$\SU(4)_L\times\SU(4)_R\to$ $\SU(4)_V$, which includes the spontaneous breaking of
U$(1)_\mathrm{PQ}$.
The corresponding $15$ Goldstone modes including the axion are decomposed in terms of $\SU(3)_\mathrm{QCD}$ 
representations as,
\begin{align}
\label{eq:NGBs in CAM}
    \mathbf{15}=\mathbf{8}\oplus 
    {\mathbf{3}}\oplus 
     \bar{\mathbf{3}}\oplus \mathbf{1} \ .
\end{align}
The color singlet Goldstone boson corresponds to the QCD axion.
The colored Goldstone modes 
obtain masses of $\order{g_\mathrm{QCD} \Lambda}$ from the QCD radiative corrections.
The key aspect of the composite axion model is the embedding of QCD and U$(1)_\mathrm{PQ}$ 
into the $\SU(4)_L\times\SU(4)_R$ flavor symmetry in axicolor dynamics.

In the original composite axion model, the U$(1)_\mathrm{PQ}$ symmetry is imposed by hand, and therefore, the model cannot address the axion quality problem. In the following section, we will discuss an extension of the composite axion model proposed in Ref. \cite{Redi:2016esr}, where the U$(1)_\mathrm{PQ}$ symmetry emerges accidentally.

%-----------------------------------------------------
\subsection{\texorpdfstring{
$\boldsymbol{\SU(3)\times\SU(N)^{n_s}\times\SU(4)^{n_s-1}}$ 
Model}{}}
\label{sec:CAAmodel}
%-----------------------------------------------------

%-----------------------------------------------------
\subsubsection*{Gauge Group and Matter Content}
%-----------------------------------------------------
In the composite accidental axion (CAA) model in Ref.\,\cite{Redi:2016esr},
the axicolor is extended 
to a product group, $\SU(N) \to \qty[\SU(N)]^{n_s}$ $(n_s\ge 2)$.
Each $\SU(N)$ gauge sector (labeled as $\SU(N)_{\mathrm{S}i}$ ($i=1,\cdots,n_s$)) consists of four pairs of fundamental and antifundamental 
left-handed Weyl fermions.
The maximal flavor symmetry 
of each sector is 
$\SU(4)_L\times \SU(4)_R\times$ U$(1)_V$ as in the case of the original composite axion model.
The subgroups of the flavor symmetries 
are weakly gauged so that the model possess $\qty[\SU(N)]^{n_s}\times \qty[\SU(4)]^{n_s-1}\times \SU(3)$ gauge symmetry. 
The moose diagram of the model is given in Fig.\,\ref{fig:moose CAA}.

Each of $n_s$ 
weakly gauged symmetries 
corresponds to the diagonal 
subgroup of the product of the $\SU(4)_L$ flavor symmetry in the $\SU(N)_{\mathrm{S}i}$ sector and the $\SU(4)_R$ flavor symmetry in the  $\SU(N)_{\mathrm{S}(i+1)}$ sector, respectively. 
Among those $n_s$ weakly coupled gauge symmetries, $n_s-1$ of them 
are gauged as $\SU(4)$ symmetries and the other one is gauged as $\SU(3)$ symmetry.
Altogether, the model has
$\qty[\SU(N)]^{n_s}\times \qty[\SU(4)]^{n_s-1}\times \SU(3)$ gauge symmetry.
Due to the axicolor dynamics,
the weakly gauged symmetry, $\qty[\SU(4)]^{n_s-1}\times \SU(3)$,
is spontaneously broken down to the
diagonal subgroup which is identified 
with $\SU(3)_\mathrm{QCD}$.
The QCD gauge coupling is given by the matching relation,
\begin{align}
    \frac{1}{g^2_\mathrm{QCD}(\Lambda)} = 
    \frac{1}{g^2_{3\mathrm{w}}(\Lambda)} +  
     \frac{1}{g^2_{4\mathrm{w},1}(\Lambda)} + 
     \cdots +
    \frac{1}{g^2_{4\mathrm{w},n_s-1}(\Lambda)} \ ,
    \label{eq:matching condition}
\end{align}
where $g_{3\mathrm{w}}, g_{4\mathrm{w},1}, \ldots,$ and $g_{4\mathrm{w},n_s-1}$ are gauge coupling constants with respect to $\SU(3)_\mathrm{w}, \SU(4)_{\mathrm{w}1}, \ldots,$ and $\SU(4)_{\mathrm{w}, n_s-1}$ respectively.
Note that some of the gauge couplings on the right hand side can be significantly larger than $g_\mathrm{QCD}$ at the breaking scale $\Lambda$,
which corresponds to the dynamical scale of the axicolor $\SU(N)_{\mathrm{S}i}$ dynamics. 
For simplicity, we assume that the dynamical scales of $\Lambda_i$ of $\SU(N)_{\mathrm{S}i}$ are equal to $\Lambda$.

To illustrate how the model works, let us 
consider the simplest setup of the model with $n_s = 2$.
The left-handed Weyl fermions in this example are listed in Tab.\,\ref{tab:model n2}.
In the table,
``w" denotes the weak coupling.
The gauge couplings of the two strong gauge symmetry are given by
$g_{\mathrm{S}1}$, $g_{\mathrm{S}2}$,
while those of weak $\SU(3)_\mathrm{w}$ and $\SU(4)_\mathrm{w}$
are given by
$g_3$ and $g_4$, respectively. 
Note that SM quarks are charged 
under $\SU(3)_\mathrm{w}$ gauge symmetry
which are not shown in Tab.\,\ref{tab:model n2}.

\begin{table}[t]
 \begin{center}
  \begin{tabular}{|c||B|R|B|R|G|G|c|c|} 
  \hline
            & $\SU(3)_{\mathrm{w}}$ & $\SU(N)_{\mathrm{S}1}$ & $\SU(4)_{\mathrm{w}}$ & $\SU(N)_{\mathrm{S}2}$ 
            & U$(1)_\mathrm{PQ}^{(\mathrm{SSB})}$
            & U$(1)_{1}$
            & U$(1)_B$
            & U$(1)_B'$
            \\ \hline
    $\psi_{A_2}^{c}$  & $\overline{\bf{3}}$ & \bf{1} & \bf{1}& \bf{N} 
    & $1$ & $1$ & $1$ & $1$ \\
    $\psi_{A_2}$  & \bf{1} 
    & \bf{1} & \bf{1} & \bf{N}
    & $-3$ & $1$ & $-3$ & $1$ \\
    $\psi_{c}^{A_1}$  & \bf{3} 
    & $\overline{\bf{N}}$ & \bf{1} & \bf{1}
    & $1$ & $1$ & $-1$ & $-1$ \\
    $\psi^{A_1}$  & \bf{1} 
    & $\overline{\bf{N}}$ & \bf{1} & \bf{1}
    & $-3$ & $1$ & $3$ & $-1$ \\ 
    $\psi_{A_1}^{p}$  & \bf{1} 
    & \bf{N} & $\overline{\bf{4}}$ & \bf{1}
    & $0$ & $-1$ & $0$ & $1$ \\
    $\psi_{p}^{A_2}$  & \bf{1} 
    & \bf{1} & \bf{4} & $\overline{\bf{N}}$
    & $0$ & $-1$ & $0$ & $-1$ \\
    \hline
  \end{tabular}
  \caption{
The minimal model of the CAA with $n_s = 2$.
The subscript ``S" denotes 
the strong coupling while ``w" denotes the weak coupling.
The indices $A_1$ and $A_2$ of the left-handed Weyl fermions $\psi$'s are for the representations under $\SU(N)_{\mathrm{S}1}$ 
and  $\SU(N)_{\mathrm{S}2}$.
The indices $c$ and $p$ are for
the representations under $\SU(3)_{\mathrm{w}}$ 
and  $\SU(4)_{\mathrm{w}}$.
The U$(1)_\mathrm{PQ}$ symmetry 
is anomalous with respect to $\SU(3)_\mathrm{w}$,
and the U$(1)_1$ symmetry is anomalous 
with respect to $\SU(3)_\mathrm{w}\times \SU(4)_\mathrm{w}$.
The global U$(1)_B$ and U$(1)_B'$ symmetries are anomaly-free 
with respect to all the gauge groups.
Blue and red columns highlight weakly and strongly coupled gauge interactions. 
The green columns highlight anomalous U(1) symmetries 
with respect to weakly coupled gauge groups.
}
  \label{tab:model n2}
 \end{center} 
\end{table}

In the limit of vanishing weak gauge couplings, $g_{3}, g_{4}\to 0$,
the global symmetry is enhanced to the
maximal symmetry
$\SU(4)_L \times \SU(4)_R \times \mathrm{U}(1)_V$ in each $\SU(N)$ gauge sector.
Hereafter, we call them 
$\SU(4)_{Li} \times \SU(4)_{Ri} \times \mathrm{U}(1)_{Vi}$ ($i=1,2$), respectively.
For $g_3\neq 0$  and 
$g_4\neq 0$,
the flavor symmetries are reduced to 
four U(1) symmetries.
Two of them are U$(1)_{V1}$ and U$(1)_{V2}$. 
In the Tab.\,\ref{tab:model n2},
we rearranged U$(1)_{V1}$ and U$(1)_{V2}$ into U$(1)_1$ and U$(1)_{B}'$,
where U$(1)_1$ is anomalous with respect to $\SU(3)_\mathrm{w}\times \SU(4)_\mathrm{w}$, while U$(1)_B'$ is free from anomaly.
The other two 
are the vector/axial combinations of 
U$(1)$ symmetries in 
the $\SU(4)_{L2}\times \SU(4)_{R1}$
flavor symmetry which commute with $\SU(3)_\mathrm{w}$. 
The axial combination is anomalous with respect to $\SU(3)_\mathrm{w}$ and is identified with U$(1)_\mathrm{PQ}$. 
The vector combination U$(1)_B$ is anomaly-free.

Note that all the U$(1)$ symmetries are
realized as accidental ones at the renormalizable level. Mass terms are also prohibited by the same reason.%
\footnote{We can also consider a more generic weakly coupled gauge group $\SU(3)_\mathrm{w}\times \qty[\SU(m)_\mathrm{w}]^{n_s-1}$.
To forbid PQ-breaking
mass terms of the fermions, we need $m\ge4$.
}
Let us also comment on the $\theta$--angles of each gauge groups denoted by
$\theta_{\mathrm{S}1}$, $\theta_{\mathrm{S}2}$, $\theta_{\mathrm{w}3}$
and $\theta_{\mathrm{w}4}$.
As for $\theta_{\mathrm{S}1}$ and $\theta_{\mathrm{S}2}$,
they can be set to zero by two global U(1) rotations
which are anomalous with respect $\SU(N)_{\mathrm{S}1}$ and $\SU(N)_{\mathrm{S}2}$.
The angles $\theta_{\mathrm{w}3}$ and $\theta_{\mathrm{w}4}$
can be set to zero by using U(1)$_\mathrm{PQ}$ and 
U(1)$_1$ rotations.
Therefore, the $\theta$--angles in this model do not spoil the axion mechanism.
These arguments can be easily extended for $n_s > 2$.
For example  we show the model contents for $n_s=3$ in Tab.\,\ref{tab:model n3}.

\begin{table}[t]
 \begin{center}
 {\scriptsize
  \begin{tabular}{|c||B|R|B|R|B|R|G|G|c|c|G|} 
  \hline
            & $\SU(3)_{\mathrm{w}}$ & $\SU(N)_{\mathrm{S}1}$ & $\SU(4)_{\mathrm{w}1}$ 
            &
            $\SU(N)_{\mathrm{S}2}$ & $\SU(4)_{\mathrm{w}2}$ & $\SU(N)_{\mathrm{S}3}$
            &              U$(1)_\mathrm{PQ}^{(\mathrm{SSB})}$
            & U$(1)_{1}$
            & U$(1)_B$
            & U$(1)_B'$
            & U$(1)_2$
            \\\hline
    $\psi_{A_3}^{c}$  & $\overline{\bf{3}}$ & \bf{1} & \bf{1} & \bf{1} & \bf{1} & \bf{N}
    & $1$ & $1$ & $1$ & $1$&$0$ \\
    $\psi_{A_3}$  & \bf{1} 
    & \bf{1} & \bf{1} & \bf{1} & \bf{1} & \bf{N}
    & $-3$ & $1$ & $-3$ & $1$&$0$ \\
    $\psi_{c}^{A_1}$  & \bf{3} 
    & $\overline{\bf{N}}$ & \bf{1} & \bf{1} & \bf{1} & \bf{1}
    & $1$ & $1$ & $-1$ & $-1$ &$0$\\
    $\psi^{A_1}$ & \bf{1} 
    & $\overline{\bf{N}}$ & \bf{1} & \bf{1} & \bf{1} & \bf{1}
    & $-3$ & $1$ & $3$ & $-1$ &$0$\\ 
    $\psi_{A_1}^{p_1}$ & \bf{1} 
    & \bf{N} & $\overline{\bf{4}}$ & \bf{1} & \bf{1} & \bf{1}
    & $0$ & $-1$ & $0$ & $1$ &$0$\\
    $\psi_{p_1}^{A_2}$ & \bf{1} 
    & \bf{1} & \bf{4} & $\overline{\bf{N}}$ & \bf{1} & \bf{1}
    & $0$ & $0$ & $0$ & $-1$ & $-1$\\
    $\psi^{p_2}_{A_2}$& \bf{1} & \bf{1} & \bf{1} & \bf{N} & $\overline{\bf{4}}$ & \bf{1}
    & $0$ & $0$ & $0$ & $1$ & $1$
   \\
   $\psi_{p_2}^{A_3}$ & $\bf{1}$& $\bf{1}$& $\bf{1}$ & $\bf{1}$ & ${\bf{4}}$ & $\overline{\bf{N}}$ 
   & $0$ & $-1$ &$0$& $-1$ & $0$
   \\ 
    \hline
  \end{tabular}
  }
  \caption{
The CAA model with $n_s=3$.
This model possesses an extra U$(1)$ symmetry U(1)$_2$, in addition to those in the model with $n_s=2$. 
Since U(1)$_2$ is anomalous only with respect to $\SU(4)_{\mathrm{w}2}$ and not broken spontaneously,
it can be used to cancel the $\theta$--angle 
of $\SU(4)_{\mathrm{w}2}$.
Meanings of the colors are the same with those in Tab.\,\ref{tab:model n2}.
}
  \label{tab:model n3}
 \end{center} 
\end{table}

%-----------------------------------------------------
\subsubsection*{Chiral Flavor Symmetry Breaking and Composite Axion}
%-----------------------------------------------------
The dynamics of the CAA model is as follows. We assume that the strong gauge interactions of $\SU(N)_{\mathrm{S}i}$ ($i=1,2$)  exhibit confinement and chiral condensations
at dynamical scales $\Lambda_i$. The VEVs of fermion bilinears are assumed to be
\begin{align}
    \langle \psi^p_{A_1}\psi^{A_1}_{\tilde{p}} \rangle&\sim \Lambda_1^3 \, \delta^{p}{}_{\tilde{p}}\ , \cr 
     \langle \psi^{\tilde{p}}_{A_2} \psi_p^{A_2} \rangle&\sim \Lambda_2^3 \, \delta^{\tilde{p}}{}_{p}\ ,
    \label{eq:condensations}
\end{align}
where $\SU(3)_\mathrm{w}$--colored $\psi^c_{A_2}$ and $\SU(3)_\mathrm{w}$--singlet $\psi_{A_2}$ are grouped together as 
$\psi^{\tilde{p}}_{A_2} = (\psi^c_{A_2},\psi_{A_2})$
and $\psi_c^{A_1},\psi^{A_1}$ are also grouped together as
$\psi_{\tilde{p}}^{A_1} = (\psi_c^{A_1},\psi^{A_1})$.
Hereafter, we take $\Lambda_1 = \Lambda_2 = \Lambda$ for simplicity.
The condensations in Eq.\,\eqref{eq:condensations}
spontaneously break $\SU(3)_{\mathrm{w}} \times \SU(4)_{\mathrm{w}}$ into $\SU(3)$ which 
is identified with $\SU(3)_\mathrm{QCD}$.
The condensations also break
U$(1)_\mathrm{PQ}$  spontaneously, while the other U$(1)$ symmetries in Tab.\,\ref{tab:model n2} are not broken spontaneously by the condensations.%
\footnote{
As for U$(1)_B$, a linear combination of U$(1)_B$ and a subgroup of $\SU(4)_{\mathrm{w}}$ generated by the 
$\mathrm{diag}(1,1,1,-3)$
remains unbroken.
}

As a result of chiral condensations, the axion 
associated with U$(1)_\mathrm{PQ}$ breaking 
appears as a composite state,
\begin{align}
\label{eq:NGBs}
\psi^p_{A_1} \psi^{A_1}_c  &\sim \Lambda^3
     \, \delta^{p}{}_{c} \, 
    e^{i\frac{a}{F_a}}\
    \ , \quad
     \psi^p_{A_1} \psi^{A_1}  
\sim   \Lambda^3 \, \delta^{p}{}_4
e^{-3i\frac{a}{F_a}}\,,
\cr
\psi^c_{A_2} \psi_p^{A_2} 
    &\sim \Lambda^3  \, 
    \delta^{c}{}_{p} \, e^{i\frac{a}{F_a}}\ , \quad
     \psi_{A_2} \psi_p^{A_2}  
\sim   \Lambda^3 \delta^{4}{}_{p}\, e^{-3i\frac{a}{F_a}} \  .
\end{align}
The domain of the axion is given by 
    ${a}/{F_a} \in [0, 2\pi)$
with $F_a$ being the axion decay constant of $\order{\Lambda}$.
The U$(1)_\mathrm{PQ}$ symmetry is anomalous under QCD, and hence, $a$ plays a role of the QCD axion.

Let us comment on the other Goldstone bosons associated with the chiral symmetry breaking, $\SU(4)_{Li}\times \SU(4)_{Ri}\to\SU(4)_{Vi}\,(i=1,2)$.
In the limit of $g_{3}, g_{4}\to0$, there are $2\times 15$ goldstone modes.
The $15$ modes of them are absorbed by the Higgs mechanism associated with the spontaneous gauge symmetry breaking,
$\SU(3)_\mathrm{w}\times \SU(4)_\mathrm{w} \to \SU(3)_\mathrm{QCD}$.
The other $15$ modes including the axion are massless at the tree-level. 
Except for the axion, they obtain masses due to the radiative corrections from $\SU(3)_\mathrm{w}\times \SU(4)_\mathrm{w}$ gauge  interactions, as in the case of the original composite axion model (see Eq.\,\eqref{eq:NGBs in CAM}).
As a result, 
all the composite states made by the axicolor dynamics obtain large masses of $\order{\Lambda}$, except for the axion.
 
%-----------------------------------------------------
\subsubsection*{Axion Quality Problem}
%-----------------------------------------------------
The advantage of the CAA model is that the PQ symmetry appears as an accidental symmetry at the renormalizable level.
This accidental symmetry is, however, 
explicitly broken by
non-renormalizable interactions.
For $n_s=2$,
the lowest-dimensional operators which violate the PQ symmetry are
\begin{align}
\mathcal{L}_{\cancel{\mathrm{PQ} } }
\sim \frac{\kappa}{M_\mathrm{Pl}^2}
\psi_c^{A_1} \psi^p_{A_1}
\psi_p^{A_2} \psi^c_{A_2}
 + \mathrm{h.c.} \ ,
\end{align}
where $M_\mathrm{Pl}$ denotes the reduced 
Planck scale and $\kappa$ is a numerical coefficient.
This should be compared with the original composite axion model, where the mass terms of the fermions, which are allowed by any gauge symmetry
can break the PQ symmetry explicitly.

The above argument can be extended straightforwardly for $n_s\ge3$, where the lowest dimensional 
PQ symmetry breaking operators are%
\begin{align}
\mathcal{L}_{\cancel{\mathrm{PQ} } }
\sim \frac{\kappa}{M_\mathrm{Pl}^{3n_s-4}}
\psi_c^{A_1}\psi^{p_1}_{A_1}
\cdots
\psi_{p_{n_s-1}}^{A_{n_s}}
\psi^c_{A_{n_s}} 
 + \mathrm{h.c.} \ .
\end{align}
These PQ breaking terms result in an additional axion potential
\begin{align}
 V_{\cancel{\mathrm{PQ}}} \sim  |\kappa|\frac{\Lambda^{3n_s} }{M_\mathrm{Pl}^{3n_s-4}}
 e^{2i a/F_a + \arg(\kappa) }
 + \mathrm{h.c}
 \ ,
\end{align}
where we have assumed that the dynamical scales of all the strong $\SU(N)$ sectors are comparable and of $\order{\Lambda}$, for simplicity.
As a result, the VEV of the axion is shifted from zero to 
\begin{align}
    \left\langle\frac{a}{F_a}\right\rangle
    \sim
    |\kappa|
    \qty(\frac{\Lambda}{M_\mathrm{Pl}})^{3n_s}
    \qty(\frac{M_\mathrm{Pl}}{\Lambda_\mathrm{QCD}})^4
    \arg(\kappa)
\end{align}
where we have used a rough estimate of the low-energy QCD contribution, $V\sim\Lambda_\mathrm{QCD}^4 \cos(2Na/F_a)$.
The axion quality problem is solved for $n_s \ge 4$, where the shift of the QCD $\theta$--angle is well below the current constraints for $\Lambda\sim10^{10}$\,GeV.
There are other possible PQ breaking operators which consist of baryonic composite operators of $\SU(N)$. Contributions from these operators are suppressed for sufficiently large $N$.

%-----------------------------------------------------
\section{Small Instanton Effects vs Chiral Symmetry
}
\label{sec:SIE vs fermions}
%-----------------------------------------------------
The CAA model in the previous section 
involves multiple broken gauge symmetries, and hence, constrained instantons potentially enhance the axion mass as we have seen in Sec.\,\ref{sec:without fermion SIE}.
In the CAA model,
however, there are new fermions.
In general, chiral symmetries of the fermions coupling to the axion have significant impacts on the axion mass.

In this section, we discuss the 
effect of chiral symmetries of the new fermions on the axion mass by considering 
a perturbative example which has the same breaking pattern of the weakly coupled gauge symmetries as well as the same chiral symmetries with the CAA model.
We will discuss the axion mass in the CAA model in Sec.\,\ref{sec:SIE in CAA}.

%-----------------------------------------------------
\subsection{Toy Example: \texorpdfstring{$\boldsymbol{\SU(3)_\mathrm{w} \times \SU(4)_\mathrm{w}}$}{} Model}
%-----------------------------------------------------
In the CAA model,
the spontaneous breaking of the 
product gauge group,
$\SU(3)_\mathrm{w} \times [\SU(4)_\mathrm{w}]^{n_s-1}$
is caused by the strong dynamics.
Here, instead, we consider a
model with gauge group $\SU(3)_\mathrm{w} \times \SU(4)_\mathrm{w}$ which is broken by condensations 
of complex scalar fields
\begin{align}
    \Phi_{\tilde{p}}{}^p\,  (\mathbf{3}\oplus \mathbf{1}, \bar{\mathbf{4}}) \ ,
    \quad
     \bar{\Phi}^{\tilde{p}}{}_p\,
     (\bar{\mathbf{3}}\oplus\mathbf{1}, \mathbf{4}) \ ,
\end{align}
with the $\SU(3)_\mathrm{w}\times \SU(4)_\mathrm{w}$ representations in the parentheses.
Here, $p$ denotes an index of the $\SU(4)_\mathrm{w}$
fundamental representation,
and $\tilde{p}$
runs the $\SU(3)_\mathrm{w}$ color ($\tilde{p}=c=1,2,3$) 
and the $\SU(3)_\mathrm{w}$ singlet ($\tilde{p}=4$).
This model mimics the CAA model with $n_s=2$, by 
assuming that the VEVs of the scalars are given by,
\begin{align}
    \langle
     \Phi_{\tilde{p}}{}^p
    \rangle = v\delta_{\tilde{p}}{}^p\ , 
    \quad
        \langle
     \bar{\Phi}^{\tilde{p}}{}_p
    \rangle = v\delta^{\tilde{p}}{}_p \ .
    \label{eq:scalar vev}
\end{align}
These VEVs break  
$\SU(3)_\mathrm{w}\times \SU(4)_\mathrm{w}$
into the diagonal subgroup $\SU(3)$
 as in the case of the CAA model.
The unbroken subgroup is identified 
with $\SU(3)_\mathrm{QCD}$.

In this toy model, 
we restrict the scalar potential 
and its coupling to the fermions as 
\begin{align}
\label{eq:toy model scalar potential} 
V & = -m^2 \Phi^\dagger \Phi 
+ \lambda (\Phi^\dagger \Phi)^2
+ \xi \det \Phi 
+ (\Phi \to \bar{\Phi})
- \kappa |\Phi\bar{\Phi}|^2
\ ,\\
\mathcal{L}_\mathrm{int}
&=  -  y_\Phi \bar{\psi}_1 \Phi \psi_2 - y_{\bar{\Phi}} \bar{\psi}_2 \bar{\Phi} \psi_1
    +\mathrm{h.c.}\ ,
    \label{eq:toy model yukawa}
\end{align}
where $\lambda$, $\xi$,  $\kappa_\Phi$ and $y_{\Phi,\bar{\Phi}}$ are coupling constants, and the mass parameter is $m^2=\order{v^2}$.
We take all the parameters real positive valued.
All the global U(1) symmetries 
of the toy model 
are listed in Tab.\,\ref{tab:model scalar},
which are the same with those in the CAA model in Tab.\,\ref{tab:model n2}.
Note that, unlike the CAA model, these symmetries are imposed by hand in Eqs.\,\eqref{eq:toy model scalar potential} and \eqref{eq:toy model yukawa}.

\begin{table}[t]
 \begin{center}
  \begin{tabular}{|c||B|B|G|G|c|c|} 
  \hline
            & $\SU(3)_{\mathrm{w}}$ 
            & $\SU(4)_{\mathrm{w}}$ 
            & U$(1)_\mathrm{PQ}^{(\mathrm{SSB})}$
            & U$(1)_{1}$
            & U$(1)_B$
            & U$(1)'_B$
            \\ \hline
    $\Phi_{\tilde{p}}{}^p$&$\bf{3}\oplus\bf{1}$&$\overline{\bf{4}}$
    &$(-1,3)$&0&$(-1,3)$&0 \\
    $\bar{\Phi}^{\tilde{p}}{}_p$&$\overline{\bf{3}}\oplus\bf{1}$&\bf{4}
    &$(-1,3)$&0&$(1,-3)$&0 \\
    \hline
    $\bar{\psi}_1^{\tilde{p}}$  & $\overline{\bf{3}}\oplus \bf{1}$ & \bf{1}
    & $(1,-3)$ & $1$ & $(1,-3)$ & $1$ \\
    $\psi_{1\tilde{p}}$  & $\bf{3}\oplus \bf{1}$ 
    & \bf{1}
    & $(1,-3)$ & $1$ & $(-1,3)$ & $-1$ \\ 
    $\bar{\psi}_2^{p}$  & \bf{1} 
    & $\overline{\bf{4}}$
    & $0$ & $-1$ & $0$ & $1$ \\
    $\psi_{2p}$  & \bf{1} 
    & \bf{4} 
    & $0$ & $-1$ & $0$ & $-1$ \\
    \hline
  \end{tabular}
  \caption{
The toy model which mimics
the CAA model with $n_s = 2$, where 
the gauge symmetry is broken by the 
VEVs of the scalar fields, ${\Phi}$ and $\bar{\Phi}$.
Meanings of colors are similar to those in the CAA model.
}
  \label{tab:model scalar}
 \end{center} 
\end{table}

As in the case of the CAA model,
the U$(1)_\mathrm{PQ}$ symmetry
and the weakly coupled $\SU(3)_\mathrm{w}\times \SU(4)_{\mathrm{w}}$ symmetries
are spontaneously broken by 
the VEVs in Eq.\,\eqref{eq:scalar vev}
simultaneously.
The corresponding axion appears in the phases,
\begin{align}
\label{eq:NGBs in scalar model}
     \Phi_c{}^{p}
    &= v  \, 
    \delta_c{}^{p} \, e^{-i\frac{a}{F_a}}\ , \quad
     \Phi_4{}^{p}
=   v\, \delta_4{}^{p}\, e^{3i\frac{a}{F_a}} \  , \cr
     \bar{\Phi}^c{}_{p}
       &= v
     \, \delta^c{}_{p} \, 
    e^{-i\frac{a}{F_a}}\ , \quad
     \bar{\Phi}^4{}_{p}
=  v \, \delta^4{}_{p}\,
e^{3i\frac{a}{F_a}}\ .
\end{align}
The axion decay constant is $F_a = \order{v}$.
As in the case of the CAA model, U$(1)$ symmetries remain unbroken except for U$(1)_\mathrm{PQ}$.

Note that there is no additional 
Goldstone mode other than the axion 
and the would-be Goldstone modes 
associated with the gauge symmetry breaking,
$\SU(3)_\mathrm{w}\times \SU(4)_\mathrm{w} \to \SU(3)_\mathrm{QCD}$.
Note also that all the fermions obtain masses of $\order{v}$ through the Yukawa interactions and the VEVs in Eq.\,\eqref{eq:scalar vev}.
As a result, this toy example leaves 
only the axion as a light particle as in the CAA model.

In the following discussion, 
we will neglect the effects of SM quarks to the axion potential.
As we will see later, 
those effects are irrelevant when comparing the QCD instanton contributions with those from the small instantons.

%-----------------------------------------------------
\subsection{Suppression of Instanton Effects by Anomalous Symmetry}
\label{sec:vanishing by anomalous symmetry}
%-----------------------------------------------------
Let us examine the impact of anomalous chiral symmetries on the axion mass within the $\SU(3)_\mathrm{w} \times \SU(4)_\mathrm{w}$ toy model.
The axion mass can be obtained from the vacuum amplitude with a constant axion background field $a$,
\begin{align}
    W(a)|_{m,n} = &
    \int
\mathcal{D}\bar{\psi}_1^{\dagger} 
     \mathcal{D}\bar{\psi}_1
     \mathcal{D}{\psi}_1^{\dagger} 
      \mathcal{D}{\psi}_1
\mathcal{D}\bar{\psi}_2^{\dagger} 
     \mathcal{D}\bar{\psi}_2 
     \mathcal{D}{\psi}_2^{\dagger}
     \mathcal{D}{\psi}_2
     \,e^{-S_\mathrm{E}[\psi,a]}\ .
\end{align}
Here $m$ and $n$ represent the winding numbers of $\SU(3)_\mathrm{w}$ and $\SU(4)_\mathrm{w}$ gauge field backgrounds, respectively.
See Appendix~\ref{app:notations} for notations of fermions in Euclidean space.
The axion dependence of $S_\mathrm{E}[\psi,a]$ appears through
\begin{align}
\label{eq:axion-fermion}
   \mathcal{L}_\mathrm{int}  
   =& -y_{\Phi} v e^{-i\frac{a}{F_a}} \bar{\psi}_1 \psi_2|_{\mathrm{colored}}
   -y_{\Phi} v e^{3i\frac{a}{F_a}} \bar{\psi}_1 \psi_2|_{\mathrm{singlet}}\cr
   &-y_{\bar{\Phi}} v e^{-i\frac{a}{F_a}} \bar{\psi}_2 \psi_1|_{\mathrm{colored}}
   -y_{\bar{\Phi}} v e^{3i\frac{a}{F_a}} \bar{\psi}_2 \psi_1|_{\mathrm{singlet}}\ .
\end{align}

In the broken phase, the U$(1)_\mathrm{PQ}$ transformation is realized by the phase rotations of $\psi_1$'s,
\begin{gather}
    \psi_{1c} \to \psi_{1c}'= e^{i\alpha_{\mathrm{PQ}} } \psi_{1c}\ , 
    \quad
        \psi_{14} \to \psi_{14}'=
        e^{-3i\alpha_{\mathrm{PQ}} } \psi_{14}\ , \\
    \quad
        \bar{\psi}_{1}^c \to \bar{\psi}_{1}^{c\prime}= e^{i\alpha_{\mathrm{PQ}} } \bar{\psi}_{1}^c\ , 
    \quad
        \bar{\psi}_{1}^4 \to \bar{\psi}_{1}^{4 \prime}= e^{-3i\alpha_{\mathrm{PQ}} } \bar{\psi}_{1}^4\ , 
\end{gather}
and a shift of the axion,
\begin{equation}
    \frac{a}{F_a} \to \frac{a}{F_a} + \alpha_\mathrm{PQ}\ , 
\end{equation}
which leaves $S_\mathrm{E}[\psi,a]$ invariant.
By using the U(1)$_\mathrm{PQ}$ transformation, we find that the vacuum amplitude satisfies,
\begin{align}
\label{eq:Wmna}
    W(a+\alpha_{\mathrm{PQ}} F_a)|_{m,n} 
    = e^{2mi\alpha_{\mathrm{PQ}}} W(a)|_{m,n}\ ,
\end{align}
where the phase factor appears due to the anomaly of U$(1)_\mathrm{PQ}$ with respect to $\SU(3)_\mathrm{w}$.
This confirms that the non-vanishing axion potential requires $m\neq 0$. 

However, this is not the end of the story. 
In this toy model, as well as in the CAA model, there exists another anomalous symmetry, U$(1)_1$, which does not undergo spontaneous breaking.
The U$(1)_1$ transformation is realized as
\begin{gather}
    \psi_{1\tilde{p}} \to \psi_{1\tilde{p}}'= e^{i\alpha } \psi_{1\tilde{p}}\ , 
    \quad
    \bar{\psi}_{1}^{\tilde{p}} \to \bar{\psi}_{1}^{\tilde{p} \prime}= e^{i\alpha } \bar{\psi}_{1}^{\tilde{p}}\ , \\
    \psi_{2p} \to \psi_{2p}'= e^{-i\alpha } \psi_{2p}\ , 
    \quad
    \bar{\psi}_{2}^p \to \bar{\psi}_{2}^{p \prime}= e^{-i\alpha } \bar{\psi}_{2}^p\ ,
\end{gather}
while the axion is not shifted.
Note that the present model possesses two $\theta$-angles, $\theta_3$ and $\theta_4$, corresponding to $\SU(3)_\mathrm{w}$ and $\SU(4)_\mathrm{w}$, respectively.
The anomalous $\mathrm{U}(1)_1$ symmetry eliminates $\theta_3-\theta_4$, leaving only QCD $\theta$-angle, $\theta_3+\theta_4$.

By using the U$(1)_1$ transformation, we find that the vacuum amplitude satisfies,
\begin{equation}
\label{eq:non-vanishing condition}
    W(a)|_{m,n} = e^{2i\alpha(m-n)}W(a)|_{m,n}\ .
\end{equation}
The phase factors appear from the 
U$(1)_1$ anomalies with respect to $\SU(3)_\mathrm{w}\times \SU(4)_\mathrm{w}$.
As a result, 
$W(a)|_{m,n}=0$ for $m\neq n$.
This means that small instanton configurations can contribute to the axion mass only when the winding numbers of $\SU(3)_\mathrm{w}$ and $\SU(4)_\mathrm{w}$ are identical.
We find that $\mathrm{U}(1)_1$ symmetry, which eliminates $\theta_3-\theta_4$, also restricts the small instanton effects on the axion mass.
As we will see later, the null contributions for $m\neq n$ can also be understood as the effect of fermion zero modes in the path integration (see Sec.\,\ref{sec:non-vanishing SIE in scalar model}).
Notice that this argument does not 
contradict with the non-vanishing 
axion potential from the QCD instantons
in this model, 
since the QCD instantons have the winding number satisfying $
m=n$.

It is instructive to note that 
the above discussion also stems from an
ambiguity of the U$(1)_\mathrm{PQ}$ symmetry.
For example, we may redefine the PQ charge assignment from that in Tab.\,\ref{tab:model scalar}
to 
\begin{align}
    Q_{\mathrm{U}(1)_\mathrm{PQ}}' = Q_{\mathrm{U}(1)_\mathrm{PQ}} + x Q_{\mathrm{U}(1)_1}\ ,
\end{align}
with an arbitrary factor $x$.
In this case, Eq.\,\eqref{eq:Wmna} becomes,
\begin{align}
    W(a + \alpha_\mathrm{PQ} F_a)|_{m,n} 
    = e^{2mi\alpha_\mathrm{PQ} + 2x(m-n)i
    \alpha_\mathrm{PQ}} W(a)\ ,
\end{align}
which is free from the ambiguity and can be non-vanishing, only for $m=n$.

%-----------------------------------------------------
\subsection{Fermion Zero Modes and 't~Hooft Operator}
%-----------------------------------------------------
As we have seen, small instantons 
can contribute to the axion mass 
only when the winding numbers of the 
$\SU(3)_\mathrm{w}$
and $\SU(4)_\mathrm{w}$ backgrounds 
coincide.
In the following, we estimate the axion mass from small instantons satisfying this condition.

To determine the non-vanishing effects of the instanton background, we must take into account the fermion zero modes around the instantons.
In general, the Dirac operator $\slashed{D}$ 
has normalizable zero modes in the instanton background.
In the case of the constrained instanton in the broken phase, 
it is notable that massive fermions 
also have normalized zero modes when 
they obtain masses from the gauge symmetry breaking field $\Phi$~\cite{Espinosa:1989qn}.
More closely, the kinetic operators of those massive fermions are given by,
\begin{align}
    i\slashed{D} - y\Phi(x)\ ,
\end{align}
which have zero modes around the 
constrained instanton in the broken phase
(see Appendix~\ref{app:zero mode}).
In the present toy model, there are zero modes in the fermions $\psi$'s around the constrained instanton associated with $\SU(3)_\mathrm{w} \times \SU(4)_\mathrm{w}
\to \SU(3)_\mathrm{QCD}$, even though they become massive in the vacuum.

%-----------------------------------------------------
\subsubsection*{'t~Hooft Operator}
%-----------------------------------------------------
To account for the effects of the fermion zero modes, it is useful to introduce the 't~Hooft operator~\cite{tHooft:1976rip,tHooft:1976snw}.
For example, let us consider an $\SU(N_c)$ gauge theory with $N_f$ pairs of left-handed Weyl fermions of the fundamental $(\psi_L)$ and the antifundamental representations $(\bar{\psi}_R)$.
In the following, we omit the index of the flavor.
Around an instanton, 
$\psi_L^\dagger$ and  $\bar{\psi}_R^\dagger$ have normalizable zero modes.
The effects of the path-integration over fermion zero modes can be captured by the insertion of a local operator at the position of the instanton.
Such operator is called as 't~Hooft operator, 
$\underset{N_f}{\det}\qty(\psi_L \bar{\psi}_R
)$.
That is, we approximate the instanton effects by, 
\begin{align}
    &\int
    \mathcal{D}\psi^\dagger
    \mathcal{D}\psi 
    \exp
    \left[
        -\int d^4x 
        \qty(
            {\bar{\psi}}_R i\slashed{D}_\mathrm{E}^{\mathrm{inst}}\bar{\psi}_R^\dagger
            +
            {\psi}^\dagger_L i\slashed{\bar{D}}_\mathrm{E}^{\mathrm{inst}}\psi_L
        )
    \right] \\
&\to       
    \int
    \mathcal{D}\psi^\dagger
    \mathcal{D}\psi \, 
    \left[
        \underset{N_f}{\det}\qty(\psi_L \bar{\psi}_{R})
        \exp\left[-\int d^4x 
        \qty({\bar{\psi}}_R
        i\slashed{D}_\mathrm{E}\bar{\psi}_R^\dagger
        +
        \psi^\dagger_L
        i\slashed{\bar{D}}_\mathrm{E}\psi_L)\right]
     \right]
   \ ,
\end{align}
where  $D^{\mathrm{inst}}_\mu$ 
and $D_\mu$ are the covariant derivatives with/without
the instanton background (see Appendix~\ref{app:notations}).
The determinant is taken over the $N_f$ flavors,
and the gauge and spinor indices of the fermions are contracted appropriately as in Ref.\,\cite{tHooft:1976rip,tHooft:1976snw}.
Similarly, the anti-instanton contribution is proportional to 
$\underset{N_f}{\det}\qty(\psi_{L}^{\dagger}\bar{\psi}^\dagger_R)$.

The 't~Hooft operators
are accompanied by 
the instanton factor and by the 
integration over sizes, that is~\cite{Csaki:2019vte},
\begin{align}
\label{eq:Det}
    \mathcal{C} \int \frac{d\rho}{\rho^5}
    \,\, \rho^{3N_f} 
    \qty(\frac{8\pi^2}{g^2})^{2N_c}
    e^{-S_\mathrm{E}(\rho^{-1})}
    \underset{N_f}{\det}\qty(\psi_L \bar{\psi}_R)   \ , 
    \quad S_\mathrm{E}(\rho^{-1}) = \frac{8\pi^2}{g^2(\rho^{-1})}\ .
\end{align}
Here, $g$ represents the gauge coupling constant of $\SU(N_c)$, and we have taken the renormalization scale to be the inverse of the small instanton size, $\mu= \rho^{-1}$~\cite{tHooft:1976snw}.
The coefficient $\mathcal{C}$ is a dimensionless constant which depends on $N_c$ and $N_f$.
The 't~Hooft operator reproduces the 
anomaly of the chiral U(1) symmetry in the presence of an $\SU(N_c)$ instanton.

%-----------------------------------------------------
\subsubsection*{'t~Hooft Operator in \texorpdfstring{$\boldsymbol{\SU(3)_\mathrm{w}\times\SU(4)_\mathrm{w}}$}{} Toy Model}
%-----------------------------------------------------
Let us go back to the $\SU(3)_\mathrm{w}\times\SU(4)_\mathrm{w}$ axion toy model. 
As we have already mentioned, fermions have zero modes around $\SU(3)_\mathrm{w}$ and $\SU(4)_\mathrm{w}$ constrained instantons. 
Let us first discuss fermion zero modes more closely. 
The fermion kinetic terms in Euclidean space are,
\begin{equation}
    \mqty(\psi_1^\dagger & \bar{\psi}_2 & \psi_2^\dagger & \bar{\psi}_1)
    \mqty(-i\barsigma^\mu D^{(3)}_{\mu} &-y_{\bar{\Phi}}\bar{\Phi}^\dagger&&\\-y_{\bar{\Phi}}\bar{\Phi}&i\sigma^\mu D^{(4)}_{\mu}&&\\&&-i\barsigma^{\mu}D^{(4)}_{\mu}&-y_{\Phi}\Phi^\dagger\\&&-y_{\Phi} \Phi&i\sigma^{\mu}D^{(3)} _{\mu} )
    \mqty(\psi_1\\\bar{\psi}_2^\dagger\\\psi_2\\\bar{\psi}_1^\dagger)
    \,,
\end{equation}
where $D_\mu^{(3)}$ and $D_\mu^{(4)}$ denote the $\SU(3)_{\mathrm{w}}$ and $\SU(4)_{\mathrm{w}}$ covariant derivatives, respectively. 
In the following, we take $y_\Phi=y_{\bar{\Phi}}=y$, for simplicity.

Following the discussions in Ref.\,\cite{Espinosa:1989qn}, of which the result is also summarized in Appendix~\ref{app:zero mode}, we see how fermion zero modes emerge around small instantons.
Around an $\SU(3)_\mathrm{w}$ constrained instanton with a size $\rho \ll v^{-1}$, fermions $\bar{\psi}_1^\dagger$ and $\psi_1^{\dagger}$ have zero modes accompanied by small $\order{\rho yv}$ contributions of $\psi_2$ and $\bar{\psi}_2$, respectively. 
When the instanton is at the origin, 
one of the zero mode wave functions is roughly expressed as,
\begin{align}
\label{eq:FZWF}
    \qty(
    {\psi_1^{\dagger(0)}(x)}\,,\,
    {\bar{\psi}{}_2^{(0)}(x)}
    )
    \sim    
    \qty( 
    \dfrac{\rho}{(x^2+\rho^2)^{3/2}} 
    \,,\, \dfrac{\rho yv}{x^2+\rho^2}
    )\,,
\end{align}
for $x,\rho\ll v^{-1}$, while they are decaying exponentially at $x \gtrsim v^{-1}$.
Around an anti-instanton, on the other hand, fermions which have zero modes can be obtained by exchanging ($\psi_{1},\psi_{2}$,$\bar{\psi}_{1},\bar{\psi}_{2}$) and ($\psi_{1}^\dagger,\psi_{2}^\dagger$,$\bar\psi_{1}^\dagger,\bar\psi_{2}^\dagger$), in Eq.\,\eqref{eq:FZWF}.
For $\SU(4)_\mathrm{w}$ (anti-)instantons, zero modes are obtained by exchanging  
$1\leftrightarrow 2$.

The model corresponds to the case with $N_f = 1$ for both $\SU(3)_\mathrm{w}$ and $\SU(4)_\mathrm{w}$
sectors.
Thus, the corresponding 't~Hooft operator for the zero modes around $\SU(3)_\mathrm{w}$ instanton is 
\begin{equation}
\label{eq:tHooft (1,0) instanton}
    \qty(\psi_1+\order{\rho yv}\bar{\psi}_2^\dagger)\qty(\bar{\psi}_1+\order{\rho yv}\psi_2^\dagger)\,,
\end{equation}
while the 't~Hooft operator for $\SU(4)_\mathrm{w}$ instantons is
\begin{equation}
\label{eq:tHooft (0,1) instanton}
    \qty(\psi_2+\order{\rho yv}\bar{\psi}_1^\dagger)\qty(\bar{\psi}_2+\order{\rho yv}\psi_1^\dagger)\,.
\end{equation}
The gauge and spinor indices are implicit, and the operators are accompanied by the integrations over collective coordinates in Eq.\,\eqref{eq:Det}. 
Note that zero modes in Eq.\,\eqref{eq:tHooft (1,0) instanton} and in Eq.\,\eqref{eq:tHooft (0,1) instanton} have $+1$ and $-1$ charges of $\mathrm{U}(1)_1$, respectively.
Accordingly, these 't~Hooft operators reproduce the 
anomalies of the U$(1)_\mathrm{PQ}$
and U$(1)_1$
symmetries in the presence of the instantons. 
Note that the differences of the U$(1)_\mathrm{PQ}$ charges, for example between $\psi_1$ and $\bar{\psi}_{2}^\dagger$ in Eq.\,\eqref{eq:tHooft (1,0) instanton}, are compensated by those of $\Phi$'s.
Hereafter, we omit $\order{\rho y v}$ contributions in the operators in Eqs.\,\eqref{eq:tHooft (1,0) instanton}
and \eqref{eq:tHooft (0,1) instanton}.

In the following, we denote the winding numbers 
$m$ of $\SU(3)_\mathrm{w}$ sector and 
$n$ of $\SU(4)_\mathrm{w}$ sector together by $(m,n)$.
The 't~Hooft operators for $(-1,0)$ and $(0,-1)$ instantons are obtained by exchanging $\psi$'s and $\psi^\dagger$'s in the operators in Eqs.\,\eqref{eq:tHooft (1,0) instanton} 
and \eqref{eq:tHooft (0,1) instanton}.

%-----------------------------------------------------
\subsection{Non-vanishing Small Instanton Effects in \texorpdfstring{$\boldsymbol{\SU(3)_\mathrm{w}\times\SU(4)_\mathrm{w}}$}{} model}
\label{sec:non-vanishing SIE in scalar model}
%-----------------------------------------------------

As we have discussed in Sec.\,\ref{sec:vanishing by anomalous symmetry},
the small instanton effects 
can induce the axion potential 
only when the winding numbers of 
$\SU(3)_\mathrm{w}$ and $\SU(4)_\mathrm{w}$ backgrounds 
coincide with each other.
Notice that the null contributions from the $m\neq n$ configurations,
as shown by Eq.\,\eqref{eq:non-vanishing condition}
can also be understood through Feynman diagrammatic arguments using 't~Hooft operators.
In the presence of instanton configurations, the path integral vanishes unless all the fermion zero modes around the instantons are closed up diagrammatically.
The numbers of the fermion zero modes $\psi_1, \bar{\psi}_1$
and $\psi_2, \bar{\psi}_2$
are equal to $2m$ and $2n$, respectively.
Due to $\mathrm{U}(1)_1$ symmetry, the difference between the numbers of fermion zero modes, $2m-2n$, remains unchanged by interactions other than 't~Hooft operators.
Therefore, when $m\neq n$,
some fermion zero modes cannot be closed up, causing the path integral to vanish.

The most relevant contributions 
are expected from $m=n=\pm1$ backgrounds,
and hence, let us first consider $(1,0)$ and $(0,1)$ instantons.
In this case, non-vanishing vacuum amplitude 
appears through diagrams in Fig.\,\ref{fig:twoinstanton}
where two blobs denote the 't~Hooft vertices 
associated with $\SU(3)_\mathrm{w}$ (i.e., $(1,0)$)
and $\SU(4)_\mathrm{w}$ (i.e., $(0,1)$) instantons, respectively.

\begin{figure}
\centering
\begin{tikzpicture}
  % Define nodes
  \node (SU3) at (0,0) {};
  \node (SU4) at (4,0) {};
  \node (center) at (2,0) {};
  % Draw curved lines
  \foreach \i in {1} {
    \draw[thick] (SU3) to[bend left=40*\i] (SU4);
    \draw[thick] (SU3) to[bend right=40*\i] (SU4);
  }
  % Draw dashed line and exponential expression
  \draw[dashed,thick] (2,.8) -- (center);
 \draw[dashed,thick] (2,-.8) -- (center);
  
  \draw[dashed, very thick] (2,.8) -- (3,2);
  \draw[dashed, very thick] (2,-.8) -- (3,-2);
  \node[above,yshift=12] at (3,1.5) {\footnotesize\( e^{i \frac{a}{F_a}} \)};
  \node[below,yshift=-12] at (3,-1.5) {\footnotesize\( e^{i \frac{a}{F_a}} \)};
  \node[left] at (2.1,.45) {\footnotesize{$\bar{\Phi}^\dagger$}};
   \node[left] at (2.1,-.45) {\footnotesize{${\Phi}^\dagger$}};
\node[] at (1.2,1.) {\footnotesize{${\psi}_1$}};
 \node[] at (1.2,-1.) {\footnotesize{$\bar{\psi}_1$}};

 \node[] at (2.8,1.) {\footnotesize{$\bar{\psi}_2$}};
 \node[] at (2.8,-1.) {\footnotesize{${\psi}_2$}};

 \node[right] at (2.13,0) {\footnotesize{$\order{v^2}$}};

  % Shading for groups
    \filldraw[fill=black!25] (SU3) circle (0.6cm)node{\footnotesize$\SU(3)_\mathrm{w}$};
    
    \filldraw[fill=black!25] (SU4) circle (0.6cm)node{\footnotesize$\SU(4)_\mathrm{w}$};
    \fill[black] (center) circle (0.23cm);
  
\end{tikzpicture}
\caption{Non-vanishing contributions to the vacuum amplitude. 
Two gray blobs denote  the 't~Hooft vertices 
associated with an $\SU(3)_\mathrm{w}$ and an $\SU(4)_\mathrm{w}$ instanton, respectively.
The axion dependence of the vacuum amplitude 
stems from the Yukawa interaction terms in Eq.\,\eqref{eq:toy model yukawa}. 
The black blob collectively describes
the connection between $\Phi$ and $\bar{\Phi}$ through the couplings to the gauge bosons and the scalar potentials. 
In the figure, the effects of SM quarks are neglected.
}
\label{fig:twoinstanton}
\end{figure}

The contributions from $(1,0)$ and $(0,1)$ instantons are accompanied by the integrations over the sizes
\begin{align}
    \mathcal{O}_{(1,0)}(x_3)=\int \frac{d\rho_3}{\rho_3^5}
    \,\, \rho_3^{3} 
    e^{-\frac{8\pi^2}{g_3^2(\rho_3^{-1})}}
     \psi_1(x_3) \bar{\psi}_1(x_3) \ , \quad
     \mathcal{O}_{(0,1)}(x_4)=\int \frac{d\rho_4}{\rho_4^5}
    \,\, \rho_4^{3} 
    e^{-\frac{8\pi^2}{g_4^2(\rho_4^{-1})}}
     \psi_2(x_4) \bar{\psi}_2(x_4) \ ,
     \label{eq:'t Hooft operators with collective coordinates}
     \end{align}
respectively. 
The coordinates $x_3$ and $x_4$ are the positions of the instantons. Note that the integration of these operators over the position of the instanton is dimensionless.
In the Feynman diagrams, the ultraviolet (UV) cutoff
on the momentum which goes through a 't~Hooft vertex 
is given by the inverse of its size, i.e. $\order{\rho^{-1}}$.
The infrared (IR) cutoff on the loop momentum is, on the other hand, given by $v$, below which the zero modes are exponentially damped.
From the loop momentum 
integration, 
we find that the vacuum amplitude is roughly given by,
\begin{equation}
\label{eq:amplitude from pair instantons}
W(a)|_{m=n=1} 
\sim    VT  \times \int\frac{\mathrm{d}\rho_3}{\rho_3^5}
    \frac{\mathrm{d}\rho_4}{\rho_4^5}
    \rho_3^3\rho_4^3v^{2}
    e^{-\frac{8\pi^2}{g^2_3(\rho_3^{-1})}}e^{-\frac{8\pi^2}{g^2_4(\rho_4^{-1})}}e^{2i\frac{a}{F_a}}\ ,
\end{equation}
where $VT$ denotes the spacetime volume.

Now, let us consider the integrations over the instanton sizes $\rho_3,\rho_4\lesssim v^{-1}$.
To extract the size dependence of the classical actions, 
we rewrite the running couplings by using the one-loop coefficients of $\beta$-functions, $b_3$ and $b_4$ for $\SU(3)_\mathrm{w}$ and $\SU(4)_\mathrm{w}$,
\begin{align}
\label{eq:RGE}
    \frac{8\pi^2}{g_{i}^2 (\rho_{i}^{-1})} 
     =  \frac{8\pi^2}{g_{i}^2 (v)} 
     + b_{i} \log \frac{1}{\rho_{i}v} \ , \quad (i = 3,4)\ .
\end{align}
By substituting these expressions, the vacuum amplitude is given by
\begin{equation}
    W(a)|_{m=n=1}\sim VT\times v^4
    e^{-\frac{8\pi^2}{g_3^2(v)}}
    e^{-\frac{8\pi^2}{g_4^2(v)}}
    e^{2i\frac{a}{F_a}}
    \int^{v^{-1}}\frac{\mathrm{d}\rho_3}{\rho_3}\qty(\rho_3v)^{b_3-1}
    \int^{v^{-1}}\frac{\mathrm{d}\rho_4}{\rho_4}\qty(\rho_4v)^{b_4-1}\,.
\end{equation}
These integrals are dominated by the
contributions from 
instantons with sizes $\rho_3,\rho_4\sim v^{-1}$ for $b_3,b_4 > 1$.
As a result, the vacuum amplitude is reduced to,
\begin{align}
    W(a)|_{m=n=1}
    &\sim
    VT\times
    v^4
    e^{-\frac{8\pi^2}{g_\mathrm{QCD}^2(v)}}e^{2i\frac{a}{F_a}}\, ,
\end{align}
for $b_3,b_4 > 1$.
Here, we have used the matching condition,
\begin{align}
\label{eq:gauge couplings in scalar model}
       \frac{1}{g_\mathrm{QCD}^2(v)}=\frac{1}{g_3^2(v)}+\frac{1}{g_4^2(v)}\ ,
\end{align}
at the symmetry breaking scale.

By adding contributions from a pair of anti-instantons,
we find that the amplitude in the small constrained instanton backgrounds for $m=n=\pm 1$ ends up with,
\begin{align}
\label{eq:axion potential in scalar model}
    W(a)|_{m=n=1}+W(a)|_{m=n=-1}
    &\sim
    VT\times
    v^4e^{-\frac{8\pi^2}{g_\mathrm{QCD}^2(v)}}\cos\left(\frac{2a}{F_a}\right)
    \,.
\end{align}
Since $e^{-8\pi^2/g_{\mathrm{QCD}}^2(v)}$ 
is small, we can use the dilute gas approximation, and we obtain the axion mass,
\begin{align}
\label{eq:SIE on axion potential}
V(a)|_\mathrm{small\,instantons}\sim v^4e^{-\frac{8\pi^2}{g_\mathrm{QCD}^2(v)}}
\cos\left(\frac{2a}{F_a}\right)
    \, ,
\end{align}
for $b_3, b_4 >1$.
Similarly, for $m=n$ with $|n|\geq2$, the instanton effects are proportional to $e^{-|n|\times8\pi^2/g^2_\mathrm{QCD}(v)}$. 
Consequently, these effects are much more suppressed compared to the contributions with $m=n = \pm 1$.

Note that in the present model, the small instanton effects cannot enhance the axion potential, even if one of the $\SU(3)_\mathrm{w}$ or $\SU(4)_\mathrm{w}$ couplings is significantly greater than the QCD coupling. 
This contrasts with the axion model without fermions discussed in Sec.\,\ref{sec:without fermion SIE}. 
In that scenario, small instantons in each $\SU(3)$ sector, such as the $(\pm1,0)$ and $(0,\pm1)$ instantons, can independently contribute to the axion potential as described in Eq.\,\eqref{eq:enhancement without fermion}. 
Consequently, the instanton effects can enhance the axion mass if one of the gauge couplings in the two sectors is much larger than the QCD coupling.

For the present model with fermions, small instantons with $m \neq n$ do not contribute to the axion potential because the fermion zero modes around the small instantons cannot be closed due to the chiral U(1)$_1$ symmetry in Tab.\,\ref{tab:model scalar}. 
To close up the fermion zero modes, non-vanishing contributions require $m=n$ small instantons in a connected diagram, as shown in Fig.\,\ref{fig:twoinstanton}.
As a result, $n=m$ small instantons contribute collectively to the axion potential, and hence, the small instantons from the larger coupling sector are accompanied by weakly-coupled small instantons from the other sector. 
Due to the matching condition in Eq.\,\eqref{eq:matching condition}, when one of the couplings is large, the other must be as small as the QCD coupling. 
Consequently, the collective instanton effects cannot enhance the axion potential.

\begin{figure}
\centering
\begin{tikzpicture}
  % Define nodes
  \node (center) at (0,0) {};
  \node (yukawa1) at (0,2.08) {};
  \node (yukawa2) at (0,-2.08) {};
  % Draw curved lines
  \foreach \i in {1} {
    \draw[thick] (center) to[bend left=22*\i] (yukawa1);
    \draw[thick] (center) to[bend right=22*\i] (yukawa1);
   \draw[thick] (center) to[bend left=22*\i] (yukawa2);
    \draw[thick] (center) to[bend right=22*\i] (yukawa2);
  }

  % Draw dashed line and exponential expression
  \draw[dashed, very thick] (yukawa1) -- (2,2.3);
  \node[right] at (2,2.3) {\footnotesize\( e^{i \frac{a}{F_a}} \)};
  
  \draw[dashed, very thick] (yukawa2) -- (2,-2.3);
  \node[right] at (2,-2.3) {\footnotesize\( e^{i \frac{a}{F_a}} \)};

  \node[] at (-0.5,1.7) {\footnotesize{$\psi_1$}};
  \node[] at (0.5,1.7) {\footnotesize{$\bar{\psi}_2$}};
  \node[] at (-0.5,-1.7) {\footnotesize{$\bar{\psi}_1$}};
  \node[] at (0.5,-1.7) {\footnotesize{$\psi_2$}};

  \node[above] at (yukawa1) {\scriptsize{$y_{\bar{\Phi}} v$}};
  \node[below] at (yukawa2) {\scriptsize{$y_\Phi v$}};

  % Shading for groups
    \fill[black] (0,2) circle (0.1cm);
    \fill[black] (0,-2) circle (0.1cm);
    \filldraw[fill=black!25] (center) circle (0.6cm)node{\scriptsize{QCD}};
\end{tikzpicture}
\caption{QCD instanton contribution to the vacuum amplitude with
size of $\rho_{\mathrm{QCD}}=\order{v^{-1}}$. 
A gray blob denotes  the 't~Hooft vertex
associated with QCD instanton. 
The axion dependence of the vacuum amplitude 
stems from the Yukawa interaction terms in Eq.\,\eqref{eq:toy model yukawa} with Eq.\,\eqref{eq:NGBs in scalar model}.
In the figure, the effects of SM quarks are neglected. 
}
        \label{fig:twoinstanton2}
\end{figure}

Let us next consider the QCD (anti-)instanton effects of the similar instanton size, i.e., $\rho_{\mathrm{QCD}}\sim v^{-1}$.
Since QCD gauge group is realized as a diagonal subgroup of $\SU(3)_\mathrm{w}$ and $\SU(4)_\mathrm{w}$,
QCD (anti-) instantons have the winding number $(m,m)$ and automatically satisfy the non-vanishing condition of the vacuum amplitude in Eq.\,\eqref{eq:non-vanishing condition}.
Accordingly, for $\rho_\mathrm{QCD} \sim v^{-1}$, 
the dominant contribution from the QCD (anti-)instanton is from $m=1$, that is,
\begin{align}
\label{eq:QCD instanton effect}
    V(a)|_{\rho_{\mathrm{QCD}}\sim v^{-1}}\, 
    \sim v^4 e^{-\frac{8\pi^2}{g^2_\mathrm{QCD}(v)}} 
    \cos\left(\frac{2a}{F_a}\right)\ .
\end{align}
A diagram for a QCD instanton contributing to the axion mass is shown in Fig.\,\ref{fig:twoinstanton2}.
The axion dependence appears 
from insertions of Eq.\,\eqref{eq:NGBs in scalar model}.
This should be compared with Fig.\,\ref{fig:twoinstanton}, where we needed a scalar loop to obtain non-vanishing contributions.
The difference stems from the fact that the 
$\psi$'s have the zero modes
of the kinetic terms around the constrained instantons, while they do not around the the QCD instantons.

The above
QCD instanton effects with 
a size of $\rho_{\mathrm{QCD}}\sim v^{-1}$ 
can be also rewritten by using $\Lambda_\mathrm{QCD}$,
\begin{equation}
\label{eq:QCD instanton effect 2}
    V(a)|_{\rho_{\mathrm{QCD}}\sim v^{-1}}\sim\Lambda_\mathrm{QCD}^4\qty(\frac{\Lambda_\mathrm{QCD}}{v})^{b_\mathrm{QCD}-4}\cos\left(\frac{2a}{F_a}\right)
    \, ,
\end{equation}
where $b_\mathrm{QCD}=7$ is the coefficient of the beta function 
of the QCD below the symmetry breaking scale $v$. 
Therefore, from Eqs.\,\eqref{eq:SIE on axion potential}, \eqref{eq:QCD instanton effect} and \eqref{eq:QCD instanton effect 2}, we find,
\begin{align}
\label{eq:comparison}
    V_{\mathrm{QCD}} \gg  V(a)|_{\rho_{\mathrm{QCD}}\sim v^{-1}}
    \sim V(a)|_\mathrm{small\,instantons}\, ,
\end{align}
where $V_\mathrm{QCD}\sim\Lambda_\mathrm{QCD}^4 \cos(2a/F_a)$.
Note that the second equality is valid for $b_3, b_4 >1 $.
As a result, we find that 
the small instanton effects do not enhance the axion potential unlike in Sec.\,\ref{sec:without fermion SIE}.

\begin{figure}
\centering
%----------------------------------
\begin{minipage}[b]{0.45\linewidth}
\centering
\begin{tikzpicture}{scale=0.95}
  \node (SU3) at (0,0) {};
  \node (SU4) at (4,0) {};
  \node (center) at (2,0) {};
  % Draw curved lines
  \foreach \i in {1} {
    \draw[thick] (SU3) to[bend left=40*\i] (SU4);
    \draw[thick] (SU3) to[bend right=40*\i] (SU4);
  }
  % Draw dashed line and exponential expression
  \draw[dashed,thick] (2,.8) -- (center);
 \draw[dashed,thick] (2,-.8) -- (center);
  
  \draw[dashed, very thick] (2,.8) -- (3,2);
  \draw[dashed, very thick] (2,-.8) -- (3,-2);
  \node[above,yshift=12] at (3,1.5) {\footnotesize\( e^{i \frac{a}{F_a}} \)};
  \node[below,yshift=-12] at (3,-1.5) {\footnotesize\( e^{i \frac{a}{F_a}} \)};
  \node[left] at (2.1,.45) {\footnotesize{$\bar{\Phi}^\dagger$}};
   \node[left] at (2.1,-.45) {\footnotesize{${\Phi}^\dagger$}};
\node[] at (1.2,1.) {\footnotesize{${\psi}_1$}};
 \node[] at (1.2,-1.) {\footnotesize{$\bar{\psi}_1$}};

 \node[] at (2.8,1.) {\footnotesize{$\bar{\psi}_2$}};
 \node[] at (2.8,-1.) {\footnotesize{${\psi}_2$}};

    %SM yukawas
    \coordinate[label=right:\textcolor{blue}{$y_u$}] (yukawa1) at (0,2){};
    \coordinate [label=right:\textcolor{blue}{$y_t$}](yukawa6) at (0,-2){};
    \coordinate[label=left:\textcolor{blue}{$y_d$}] (yukawa2) at (90+36:2){};
    \coordinate [label=above:\textcolor{blue}{$y_s$}](yukawa3) at (90+36+36:2){};
    \coordinate [label=below:\textcolor{blue}{$y_c$}](yukawa4) at (90+36*3:2){};
    \coordinate [label=left:\textcolor{blue}{$y_b$}](yukawa5) at (90+36*4:2){};

    \fill[blue] (yukawa1) circle (0.12cm);
    \fill[blue] (yukawa2) circle (0.12cm);
    \fill[blue] (yukawa3) circle (0.12cm);
    \fill[blue] (yukawa4) circle (0.12cm);
    \fill[blue] (yukawa5) circle (0.12cm);
    \fill[blue] (yukawa6) circle (0.12cm);

    %SM quarks
    \draw[thick,blue] (SU3) to[out=90+16,in=270-16] (yukawa1);
    \draw[thick,blue] (SU3) to[bend right=16]node[xshift=4,yshift=8]{$u$} (yukawa1);
    
    \draw[thick,blue] (SU3) to[bend left=16] (yukawa2);
    \draw[thick,blue] (SU3) to[out=90+36-16,in=270+36+16]node[xshift=-1,yshift=10]{$d$} (yukawa2);
    
    \draw[thick,blue] (SU3) to[bend left=16] (yukawa3);
    \draw[thick,blue] (SU3) to[bend right=16]node[xshift=-5,yshift=6]{$s$} (yukawa3);
    
    \draw[thick,blue] (SU3) to[bend left=16] (yukawa4);
    \draw[thick,blue] (SU3) to[bend right=16]node[xshift=-6,yshift=2]{$c$} (yukawa4);
    
    \draw[thick,blue] (SU3) to[bend left=16] (yukawa5);
    \draw[thick,blue] (SU3) to[bend right=16]node[xshift=-8,yshift=-3]{$b$} (yukawa5);
    
    \draw[thick,blue] (SU3) to[bend left=16] (yukawa6);
    \draw[thick,blue] (SU3) to[bend right=16]node[xshift=-3,yshift=-10]{$t$} (yukawa6);

    \draw[very thick, dashed,blue](yukawa1) to[bend right=16]node[above]{$H_\mathrm{SM}$} (yukawa2);

    \draw[very thick, dashed,blue](yukawa3) to[bend right=16]node[left]{$H_\mathrm{SM}$} (yukawa4);

    \draw[very thick, dashed,blue](yukawa5) to[bend right=16]node[below]{$H_\mathrm{SM}$}(yukawa6);

    % Shading for groups
    \filldraw[fill=black!25] (SU3) circle (0.6cm)node{\footnotesize$\SU(3)_\mathrm{w}$};
    
    \filldraw[fill=black!25] (SU4) circle (0.6cm)node{\footnotesize$\SU(4)_\mathrm{w}$};
    \fill[black] (center) circle (0.23cm);
\end{tikzpicture}

\subcaption[]{\label{fig:SM zero modes in constrained instanton}Small (1,0) and (0,1) instantons}
\end{minipage}
%%%%%%%%%%%%%%%%%%%%%%%%%%%%%%%%%%%%%%%%%%%%%%%%%%
\begin{minipage}[b]{0.45\linewidth}
\centering
\begin{tikzpicture}{scale=0.95}
    % Define nodes
  \node (center) at (0,0) {};
  \coordinate (yukawa1) at (22.5:2) {};
  \coordinate [,label=right:\textcolor{blue}{$y_u$}](yukawa2) at (22.5+45:2) {};
  \coordinate  [label=left:\textcolor{blue}{$y_d$}](yukawa3) at (22.5+90:2) {};
  \coordinate [label=above:\textcolor{blue}{$y_s$}](yukawa4) at (22.5+135:2) {};
  \coordinate [label=below:\textcolor{blue}{$y_c$}](yukawa5) at (22.5+180:2) {};
  \coordinate [label=left:\textcolor{blue}{$y_b$}](yukawa6) at (22.5+225:2){};
  \coordinate [label=right:\textcolor{blue}{$y_t$}](yukawa7) at (22.5+270:2){};
  \coordinate (yukawa8) at (22.5+315:2){};
  % Draw curved lines
  % Draw new loops
    \draw[thick] (center) to[bend left=22] node[xshift=15,yshift=10]{\footnotesize$\psi_1$}(yukawa1);
    \draw[thick] (center) to[bend right=22] node[xshift=18,yshift=2]{\footnotesize$\bar{\psi}_2$}(yukawa1);
    \draw[thick,blue] (center) to[bend left=22](yukawa2);
    \draw[thick,blue] (center) to[bend right=22] node[yshift=10,xshift=9]{$u$}(yukawa2);
    \draw[thick,bend left=22,blue] (center) to (yukawa3);
    \draw[thick,bend right=22,blue] (center) to node[xshift=0,yshift=15]{$d$} (yukawa3);
    \draw[thick,blue] (center) to[bend left=22] (yukawa4);
    \draw[thick,blue] (center) to[bend right=22] node[xshift=-5,yshift=8]{$s$} (yukawa4);
    \draw[thick,blue] (center) to[bend left=22] (yukawa5);
    \draw[thick,blue] (center) to[bend right=22] node[xshift=-10,yshift=1]{$c$}(yukawa5);
    \draw[thick,blue] (center) to[bend left=22](yukawa6);
    \draw[thick,blue] (center) to[bend right=22] node[xshift=-7,yshift=-5]{$b$}(yukawa6);
    \draw[thick,blue] (center) to[bend left=22] (yukawa7);
    \draw[thick,blue] (center) to[bend right=22] node[xshift=0,yshift=-10]{$t$}(yukawa7);
    \draw[thick] (center) to[bend left=22]node[xshift=18,yshift=-3]{\footnotesize$\psi_2$} (yukawa8);
    \draw[thick] (center) to[bend right=22]node[xshift=15,yshift=-12]{\footnotesize$\bar{\psi}_1$}(yukawa8);

  % Draw dashed line and exponential expression

  \draw[dashed, very thick, blue] (yukawa2) to[bend right=22]node[above]{$H_\mathrm{SM}$}(yukawa3);
  \draw[dashed, very thick, blue] (yukawa4) to[bend right=22]node[left]{$H_\mathrm{SM}$}(yukawa5);
  \draw[dashed, very thick, blue] (yukawa6) to[bend right=22]node[below]{$H_\mathrm{SM}$} (yukawa7);
  
  \coordinate (axp) at (2.5,1);
  \coordinate (axm) at (2.5,-1);
  
  \draw[dashed, very thick] (yukawa1) -- (axp);
  \node[right] at (axp) {\footnotesize\( e^{i \frac{a}{F_a}} \)};
  \draw[dashed, very thick] (yukawa8) -- (axm);
  \node[right] at (axm) {\footnotesize\( e^{i \frac{a}{F_a}} \)};

  % Shading for groups
    \fill[black] (yukawa1) circle (0.12cm);
    \fill[blue] (yukawa2) circle (0.12cm);
    \fill[blue] (yukawa3) circle (0.12cm);
    \fill[blue] (yukawa4) circle (0.12cm);
    \fill[blue] (yukawa5) circle (0.12cm);
    \fill[blue] (yukawa6) circle (0.12cm);    
    \fill[blue] (yukawa7) circle (0.12cm);
    \fill[black] (yukawa8) circle (0.12cm);
    \filldraw[fill=black!25] (center) circle (0.6cm)node{\scriptsize{QCD}};
\end{tikzpicture}
\subcaption[]{Small QCD instanton}
\end{minipage}
%%%%%%%%%%%%%%%%%%%%%%%%%%%%%%%%%
\caption{'t~Hooft operators with SM quark zero modes closed by SM Higgs loops. 
In both figures, the instanton sizes are approximately $\rho \lesssim v^{-1}$, significantly smaller than the inverse electroweak scale. 
The SM contributions are highlighted in blue.}
\label{fig:SM zero modes}

\end{figure}

We have omitted the effects of SM quarks so far.
Here, we will discuss the effects of the quark zero modes. 
Instanton effects become non-zero only when the quark zero modes are closed by mass insertions or some interactions.
Since we are comparing the constrained instantons and small QCD instantons of sizes $\rho\sim v^{-1}$, the dominant contributions arise from closing the SM quark zero modes through SM Higgs loops, as illustrated in Fig.\,\ref{fig:SM zero modes}. 
These effects are common to both constrained instantons and QCD instantons of small scales, providing the suppression factor via SM quark Yukawa couplings,
\begin{equation}
    D_y \simeq \prod_{i=u,d,s,c,b,t}\left(\frac{y_i}{4\pi}\right)\,.
\label{eq:SM zero mode suppression}
\end{equation}
This factor is applied to the axion potential
$V(a)|_{\rho_{\mathrm{QCD}}\sim v^{-1}}$ and $V(a)|_\mathrm{small\,instantons}$.
Since this suppression is common to both QCD instanton contributions and constrained instanton contributions of similar sizes, the comparisons in 
Eq.\,\eqref{eq:comparison}
remain valid, even in the presence of SM quarks.

Two comments are in order.
First, in Eq.\,\eqref{eq:amplitude from pair instantons}, we have neglected the $\order{\rho_{3,4}^2 v^2}$ contributions 
in the classical action for the constrained instantons in Eq.\,\eqref{eq:action w/ constrained instanton}.
Since the integration over the instanton sizes is dominated by $\rho_3,\rho_4 \sim v^{-1}$, the $\order{\rho_{3,4}^2 v^2}$ contributions become sizable.
In addition, we have also neglected the effect of overlapping of two constrained instantons.
When two constrained instantons get close comparable to their respective sizes, i.e. $|x_3-x_4| \lesssim \rho_{3,4}$, 
the configurations deviate from the 
isolated constrained instanton configurations.
Thus, we expect that those effects enhance the value of the classical
action, that is,
\begin{align}
    S_\mathrm{E} > S_\mathrm{E3}(v^{-1}) + S_\mathrm{E4}(v^{-1})\ ,
\end{align}
where $S_\mathrm{E3,4}(\rho_{3,4})$ are the classical action for the isolated constrained instanton in Eq.\,\eqref{eq:action w/ constrained instanton}.
Since the vacuum amplitude is suppressed 
by $e^{-S_\mathrm{E}}$, the larger $S_\mathrm{E}$ results in
a smaller vacuum amplitude.
Therefore, the small constrained instanton contributions estimated in this section should be regarded as rough upper limits.

Second, in the above discussion, we have assumed $b_3, b_4 > 1$.
When $b_3, b_4 < 1$, on the other hand,
the instantons with sizes much smaller than $v^{-1}$ can 
be dominant and the estimate in Eq.\,\eqref{eq:SIE on axion potential} is altered. 
In the CAA model, we will see that 
the effects from instantons much smaller 
than the dynamical scale of the axicolor dynamics are more suppressed from a dynamical reason.

%-----------------------------------------------------
\section{
Small Instanton Effects on Composite Accidental Axion
}
\label{sec:SIE in CAA}
%-----------------------------------------------------
Let us examine small instanton effects in the CAA model. 
Note that global symmetries and gauge groups relevant for the discussion of the axion mass are identical between the CAA model and the toy model in the previous section.
In the following, 
we discuss the small instanton effects in the CAA model by repeating 
the arguments in the previous section.

%-----------------------------------------------------
\subsection{Vanishing Small Instanton Effects }
\label{sec:vanishing Effects}
%-----------------------------------------------------
Let us consider the $n_s = 2$ model.
We can extend the following discussion
for $n_s > 2$ straightforwardly.
As discussed in the previous section,
the effects of SM quark zero modes  
can be incorporated with small instanton effects by applying the suppression factor in Eq.\,\eqref{eq:SM zero mode suppression}.
In the following, we omit this factor, since it is irrelevant for the comparison between the effects of constrained instantons and small QCD instantons.

The axion mass can be again obtained from the vacuum amplitude with a constant axion background field $a$,
\begin{align}
\label{eq:WmnCAA}
       W(a)|_{m,n} = \int \prod \mathcal{D}A_N\mathcal{D}\psi^\dagger \mathcal{D}\psi \,e^{-S_\mathrm{E}[\psi,A_N]}  \ ,
\end{align}
where $m$ and $n$ represent the winding numbers of $\SU(3)_\mathrm{w}$ and $\SU(4)_\mathrm{w}$ gauge field backgrounds, respectively.
Here, $\psi$ and $\psi^\dagger$ collectively denote the fermions, while
$A_N$ are the gauge fields of the axicolor $\SU(N)_{\mathrm{S}i}$ dynamics.
The axion dependence of Eq.\,\eqref{eq:WmnCAA}
appears through Eqs.\,\eqref{eq:NGBs}. 
Note again that the axicolor dynamics does not break the U$(1)_\mathrm{PQ}$ symmetry, and hence, does not generate the axion potential by itself.

Now, let us consider the 
 U(1)$_1$ rotation $\psi\to\psi'$ in Tab.\,\ref{tab:model n2}, 
\begin{gather}
    \psi^{\prime \tilde{p}}_{A_2} = e^{i\alpha}\psi^{\tilde{p}}_{A_2}, \quad
    \psi^{\prime A_1}_{\tilde{p}} = e^{i\alpha}\psi^{A_1}_{\tilde{p}}\,, \\
    \psi^{\prime p}_{A_1} = e^{-i\alpha}\psi^{p}_{A_1}
    ,\quad
    \psi^{\prime A_2}_{p} = e^{-i\alpha}\psi^{A_2}_{p}
    \, . 
\end{gather}
Note that the axion is not affected by the U$(1)_1$ rotation.
Under this rotation, the vacuum amplitude changes its phase as,
 \begin{align}
    W(a)|_{m,n}
    &=
    \int \prod
    \mathcal{D}A_N
    \mathcal{D}\psi'^\dagger
    \mathcal{D}\psi'
    \,
    e^{-S_{\mathrm{E}}[\psi',A_N]}
    \cr
    &=
    \int \prod
    \mathcal{D}A_N
    \mathcal{D}\psi^\dagger
    \mathcal{D}\psi
    \,
    \exp\!\!
    \qty[
        2N
        \qty(
            \int\!\mathrm{d}^4x\frac{1}{32\pi^2}{F_3}^a_{\mu\nu}\tilde{F_3}^{a\mu\nu}
            \!\!
            -\!\!
            \int\!\mathrm{d}^4x\frac{1}{32\pi^2}{F_4}^a_{\mu\nu}\tilde{F_4}^{a\mu\nu}\!
        )
        i\alpha
    ]
    e^{-S_\mathrm{E}[\psi,A_N]}
    \cr
    &=
    e^{2N(m-n)i\alpha} W(a)|_{m,n}\,.
    \label{eq:path integral anomaly in CAA}
\end{align}
Here, $F_3$ and $F_4$ represent the field strengths of $\SU(3)_\mathrm{w}$ and $\SU(4)_\mathrm{w}$ gauge fields, respectively,
and the second equality is the result of the 
chiral anomaly of U$(1)_1$.
Therefore, as in the toy model, 
we find that the vacuum amplitude vanishes 
unless $m= n$.

%-----------------------------------------------------
\subsection{Non-vanishing Small Instanton Effects}
\label{sec:non-vanising Efects}
%-----------------------------------------------------
Let us consider the small instanton effects with $m=n$.
In particular, 
we focus on the effects from  a pair of $(\pm1,0)$ and $(0,\pm1)$ instantons, which provide 
the dominant contribution to the axion potential.

Let us consider small constrained instantons with sizes $ \rho \ll \Lambda^{-1}$, around which the fermions have zero modes of the kinetic terms.
The relevant 't~Hooft operators which encapsulate the effects of the zero modes
around 
the $(1,0)$ and $(0,1)$ instantons in the CAA model are proportional to,
\begin{align}
   \mathcal{O}_{(1,0)}= \int \frac{d\rho_3}{\rho_3^5}
    \,\, \rho_3^{3N} 
    e^{-\frac{8\pi^2}{g_{3}^2(\rho_3^{-1})}}
     \underset{A_1,A_2}{\det}({{\psi}}_{A_2}^{c'}{{\psi}}^{A_1}_{c})   
     \,, \quad
   \mathcal{O}_{(0,1)} = \int \frac{d\rho_4}{\rho_4^5}
    \,\, \rho_4^{3N}
    e^{-\frac{8\pi^2}{g_{4}^2(\rho_4^{-1})}}
    \underset{A_1,A_2}{\det}({\psi}_{A_1}^{p'}{\psi}^{A_2}_{p})\,,
     \label{eq:'t Hooft operators with collective coordinates in CAA}
\end{align}
where $\rho_{3},\rho_{4} \ll \Lambda^{-1}$
(see Eq.\,\eqref{eq:Det}).
The integration of these operators over the position of the instanton is dimensionless. 
The determinants are taken over the 
axicolor indices.

In the toy model in the previous section,
the 't Hooft operators $\mathcal{O}_{(1,0)}$ and
$\mathcal{O}_{(0,1)}$
are connected through the Yukawa interactions 
with a scalar loop (see Fig.\,\ref{fig:twoinstanton}).
In the CAA model, on the other hand, the operators
$\mathcal{O}_{(1,0)}$ and
$\mathcal{O}_{(0,1)}$ are connected through 
the effective interactions such as,
\begin{align}
\label{eq:symmetric}
{\mathcal{O}}_{\mathrm{sym}}  &\sim 
      M^{4-6N}
       [(\psi^{A_1}_c\psi_{A_1}^p)(\psi_{A_2}^c\psi^{A_2}_p)]^N
      \underset{A_1,A_2}{\det}({{\psi}{}^\dagger}^{A_2}_c{{\psi}{}^\dagger}_{A_1}^{c})
      \underset{A_1,A_2}{\det}({\psi^\dagger}^{A_1}_p{\psi^\dagger}_{A_2}^{p}) +\mathrm{h.c.}\, ,
\end{align}
which are generated by the axicolor strong dynamics.
Here, $M$ denotes the Wilsonian cutoff scale for the effective operator.
Let us emphasize that 
the operators generated by the axicolor 
dynamics do not violate the U$(1)_\mathrm{PQ}$ and U$(1)_1$ symmetries, since those U$(1)$ symmetries are anomaly-free with respect to $\SU(N)_{\mathrm{S}i}$.

\begin{figure}
\centering

\begin{tikzpicture}
  % Define nodes
  \node (SU3) at (0,0) {};
  \node (SU4) at (4,0) {};
  \node (center) at (2,0) {};
  % Draw curved lines
  \foreach \i in {.6,1.5,2.5} {
    \draw[thick] (SU3) to[bend left=25*\i] (center);
    \draw[thick] (SU3) to[bend right=25*\i] (center);
   \draw[thick] (SU4) to[bend left=25*\i] (center);
    \draw[thick] (SU4) to[bend right=25*\i] (center);
  }

  % Draw dashed line and exponential expression
  \draw[dashed,very thick] (2,1) -- (center);
  \node[above] at (2,1) {\footnotesize\( e^{i \frac{2Na}{F_a}} \)};

  % Shading for groups
    \filldraw[fill=black!25] (SU3) circle (0.6cm)node{\footnotesize$\SU(3)_\mathrm{w}$};
    
    \filldraw[fill=black!25] (SU4) circle (0.6cm)node{\footnotesize$\SU(4)_\mathrm{w}$};
    \fill[black] (center) circle (0.18cm);

    \node[]() at (1.3,.27){\scriptsize:};
    \node[]() at (4-1.3,.27){\scriptsize:};
    \node[]() at (1.3,-.27){\scriptsize:};
    \node[]() at (4-1.3,-.27){\scriptsize:};
\end{tikzpicture}

\caption{Non-vanishing contributions to the vacuum amplitude. 
Two gray blobs denote the 't~Hooft vertices 
associated with an $\SU(3)_\mathrm{w}$ and an $\SU(4)_\mathrm{w}$ instanton, respectively.
The black blob describes
the effective interaction term in Eq.\,\eqref{eq:symmetric}. 
As in Fig.\,\ref{fig:SM zero modes in constrained instanton}, 
there are SM quark zero modes around $\SU(3)_\mathrm{w}$ instanton, which are closed by SM Higgs loops. 
These are omitted in this figure.
}
\label{fig:twoinstanton in CAA}
\end{figure}

With the operator $\mathcal{O}_\mathrm{sym}$, 
the fermion zero modes in $\mathcal{O}_{(1,0)}$ and $\mathcal{O}_{(0,1)}$ are closed up as shown in Fig.\,\ref{fig:twoinstanton in CAA}.
The axion dependence appears by inserting 
Eqs.\,\eqref{eq:NGBs} to Eq.\,\eqref{eq:symmetric}.
To estimate the vacuum amplitude,
let us assume that the integration over 
the loop momenta $\ell$'s
in Fig.\,\ref{fig:twoinstanton in CAA} is dominated by 
$\ell \sim \ell_{\mathrm{dom}}$,
where $\ell_{\mathrm{dom}}$ is in between
$\Lambda \lesssim \ell_{\mathrm{dom}} \lesssim \rho_{3,4}^{-1}$.
In this case, we can crudely estimate the 
vacuum amplitude by substituting 
\begin{align}
M^{4-6N} &\sim \ell_{\mathrm{dom}}^{4-6N}\ ,\\
    [(\psi^{A_1}_c\psi_{A_1}^p)(\psi_{A_2}^c\psi^{A_2}_p)]^N &\sim 
    \Lambda^{6N} e^{2Ni \frac{a}{F_a} }\ ,
\label{eq:NGBs into symmetric operator}
\end{align}
which results in
\begin{align}
\label{eq:W11 in CAA}
  W(a)|_{m=n=1}
    &\sim
    VT\times 
    \int^{\Lambda^{-1}}\frac{\mathrm{d}\rho_3}{\rho_3^5}\rho_3^{3N}
    e^{-\frac{8\pi^2}{g_{3}^2(\rho_3^{-1})}}
    \int^{\Lambda^{-1}}\frac{\mathrm{d}\rho_4}{\rho_4^5}\rho_4^{3N}
    e^{-\frac{8\pi^2}{g_{4}^2(\rho_4^{-1})}}
    \times
    \ell_\mathrm{dom}^{-4}
    \times
    \Lambda^{6N}e^{2Ni\frac{a}{F_a}}\ .
\end{align}    
By noting $\ell_{\mathrm{dom}}^{-1} \lesssim \Lambda^{-1}$, we find that the size of the 
vacuum amplitude from the small instanton contributions is limited as,
\begin{align}
    \Big|{W(a)|_{m=n=1}}\Big|
    &\lesssim
    VT\times 
    \int^{\Lambda^{-1}}\frac{\mathrm{d}\rho_3}{\rho_3^5}\rho_3^{3N}
    e^{-\frac{8\pi^2}{g_{3}^2(\rho_3^{-1})}}
    \int^{\Lambda^{-1}}\frac{\mathrm{d}\rho_4}{\rho_4^5}\rho_4^{3N}
    e^{-\frac{8\pi^2}{g_{4}^2(\rho_4^{-1})}}
    \times
    \Lambda^{6N-4}
    \cr
    &\sim
    VT\times \Lambda^4
    e^{-\frac{8\pi^2}{g_{3}^2(\Lambda)}}
    e^{-\frac{8\pi^2}{g_{4}^2(\Lambda)}}\int^{\Lambda^{-1}}\frac{\mathrm{d}\rho_3}{\rho_3}\qty(\rho_3\Lambda)^{b_3+3N-4}
    \int^{\Lambda^{-1}}\frac{\mathrm{d}\rho_4}{\rho_4}\qty(\rho_4\Lambda)^{b_4+3N-4}.
\end{align}
Here, we have used the running gauge coupling constants in Eq.\,\eqref{eq:RGE} to obtain the final expression.
The IR cutoff of the integration
over $\rho_{3,4}$ is of $\order{\Lambda^{-1}}$.

From this expression, we find that the integration over the instanton sizes is dominated 
by the IR contribution, i.e., $\rho_{3,4}\simeq \Lambda^{-1}$, for  
\begin{align}
\label{eq:b34 low}
    b_{3,4} > 4 - 3N\ .
\end{align}
As the $\beta$-function coefficients
in the CAA model are given by,
\begin{align}
    b_3 &= b_\mathrm{QCD} - \frac{2}{3}N =  7 - \frac{2}{3} N \ , \\
    b_4 &= \frac{44}{3} - \frac{2}{3}N \  ,
\end{align}
the condition \eqref{eq:b34 low} is 
satisfied for both $\SU(3)_\mathrm{w}$ and $\SU(4)_\mathrm{w}$ with $N>0$.
Therefore, in the CAA model, the crude upper limit on the vacuum amplitude from the small constrained instanton effects is given by,
\begin{align}
    \Big|W(a)|_{m=n=1}\Big|
    &\lesssim
    VT\times
    \Lambda^4
    e^{-\frac{8\pi^2}{g_{3}^2(\Lambda)}}
    e^{-\frac{8\pi^2}{g_{4}^2(\Lambda)}}\,.
\end{align}
Accordingly, 
the axion potential 
from the small instanton effects 
is at most,
\begin{align}
\label{eq:SIE on axion potential in CAA}
V(a)|_\mathrm{small\,instantons}\sim \Lambda^4e^{-\frac{8\pi^2}{g_\mathrm{QCD}^2(\Lambda)}}
\cos\left(\frac{2Na}{F_a}\right)
    \, .
\end{align}
Here, we have used the matching condition in Eq.\,\eqref{eq:gauge couplings in scalar model} at $\Lambda$.
In the presence of SM quarks, the suppression factor in Eq.\,\eqref{eq:SM zero mode suppression} is applied to the above contribution. 
This factor is common to both constrained instantons and small QCD instantons.

We have found that the small instanton effects do not enhance the axion mass in the CAA model. 
This observation is similar to the case of the toy model discussed in Sec.\,\ref{sec:non-vanishing SIE in scalar model}. 
Specifically, small instantons with $m \neq n$ do not contribute to the axion potential due to the chiral U(1)$_1$ symmetry shown in Tab.\,\ref{tab:model n2}.
The collective $m=n$ instanton effects are suppressed by $e^{-|m|\times 8\pi^2/g^2_3(\Lambda)}\times e^{-|n|\times 8\pi^2/g^2_4(\Lambda)}$, which coincides with $e^{-|n|\times 8\pi^2/g^2_\mathrm{QCD}(\Lambda)}$ due to
the matching condition of the QCD 
coupling in Eq.\,\eqref{eq:matching condition}.

Several comments are in order.
First, in Eq.\,\eqref{eq:W11 in CAA}, 
we have used 
Eq.\,\eqref{eq:NGBs into symmetric operator} regardless of the size of the dominant loop momentum $\ell_{\mathrm{dom}}$.
This substitution overestimates the size of the vacuum amplitudes if $\ell_\mathrm{dom}\gg \Lambda$, since the condensation of the fermion bilinears disappear at the scale much higher than $\Lambda$.
Second, we have also neglected the $\order{\rho^2\Lambda^2}$ contributions 
to the classical action for the constrained instantons.
Third, we have also neglected the increase of the value of the classical action due to the overlapping of the constrained instantons.
These two effects lead to an underestimation of the classical action and therefore to an overestimation of the vacuum amplitude.
Putting 
altogether,
the small constrained instanton contributions estimated
in this section should be regarded as rough upper limits.

%-----------------------------------------------------
\subsection{Small Instanton Effects for \texorpdfstring{$\boldsymbol{n_s > 2}$}{}}
%-----------------------------------------------------
Finally, we mention the CAA models with $n_s > 2$.
In the above discussion for $n_s = 2$,
the U$(1)_1$ global symmetry plays an important role to show that only the $m=n$ instantons 
contribute to the axion potential. The crucial feature of U$(1)_1$ is that 
it is free from the axicolor anomaly and also not broken
spontaneously by the axicolor dynamics,
while being anomalous with respect to $\SU(3)_\mathrm{w} \times \SU(4)_\mathrm{w}$.
In models with $n_s >2$, we have 
$n_s -1$ global U$(1)$ symmetries, U$(1)_i$ ($i=1,\cdots, n_s-1$), which have 
the same roles of U$(1)_1$ in the model with 
$n_s = 2$.
For example, $n_s=3$ model possesses 
U$(1)_2$ 
in Tab.\,\ref{tab:model n3}, in addition to U$(1)_1$.

In the model with $n_s \ge 2$, the vacuum amplitudes are labeled by
the winding numbers of $\SU(3)_\mathrm{w}$,
$\SU(4)_{\mathrm{w},1}$, $\cdots$, $\SU(4)_{\mathrm{w}, n_s-1}$ sectors denoted by $(m,n_1,\cdots,n_{n_s-1})$.
Under the $n_s-1$ anomalous U$(1)$ transformations,
the vacuum amplitude labeled by
$(m,n_1,\cdots, n_{n_s-1})$ changes its phase as
\begin{align}
    W(a)|_{m,n_1,\cdots,n_{n_s-1}} 
    = e^{i\alpha_1 (2 m - n_1 - n_{n_s -1})}
    e^{i\alpha_2 (n_2 - n_1)}
    \cdots 
   e^{i\alpha_{n_s-1}(n_{n_s-1} - n_{n_s-2})} 
    \times W(a)|_{m,n_1,\cdots,n_{n_s-1}}\ ,
\end{align}
where $\alpha_i$ ($i = 1, \cdots , n_s - 1$) are the rotation angles of U$(1)_i$ transformations. 
Thus, we again find that the contributions from the small constrained instantons vanish unless 
\begin{align}
    m = n_1  = \cdots = n_{n_s-1}\ .
\end{align}

As a result, by repeating the discussion in Sec.\,\ref{sec:non-vanising Efects},
we find that the small instanton effects are at most
\begin{align}
V(a)|_\mathrm{small\,\,instanton } \sim
\Lambda^4 e^{-\frac{8\pi^2}{g^2_\mathrm{QCD}(\Lambda)}}\cos\left(\frac{2Na}{F_a}\right)\ .
\end{align}
Here, the QCD coupling is matched to the gauge couplings of $\SU(3)_\mathrm{w} \times [\SU(4)_\mathrm{w}]^{n_s -1}$ via
\begin{align}
    \frac{1}{g^2_\mathrm{QCD}(\Lambda)} = 
    \frac{1}{g^2_{3\mathrm{w}}(\Lambda)} +  
     \frac{1}{g^2_{4\mathrm{w},1}(\Lambda)} + 
     \cdots +
    \frac{1}{g^2_{4\mathrm{w},n_s-1}(\Lambda)}      \ ,
\end{align}
at the symmetry breaking scale $\Lambda$.
Therefore, the axion mass is not enhanced by small constrained instantons, also in the case of $n_s>2$.

%-----------------------------------------------------
\section{Application to Other  Composite Axion Models}
\label{sec:SIE in other models}
%-----------------------------------------------------
As we have learned in the previous section, the small instanton effects do not enhance the axion mass in the CAA model. 
In this section, let us extend our discussion to other types of composite axion models.

%-----------------------------------------------------
\subsection{Composite Axion Models with Spectator QCD}
%-----------------------------------------------------
Let us first consider a class of composite axion models in which $\SU(3)_\mathrm{QCD}$ does not take part in the chiral symmetry breaking caused by the axicolor dynamics. 
Specifically, we examine a model with 
gauge symmetry of $G_\mathrm{S}\times G_\mathrm{w}\times \SU(3)_\mathrm{QCD}$
where $G_\mathrm{S}$ represents the axicolor gauge group, and $G_\mathrm{w}$ is a weakly gauged subgroup of flavor symmetry. 
In these models, the axicolor dynamics only spontaneously breaks $G_\mathrm{w}$.
QCD, therefore, acts as a spectator to the axicolor dynamics. 
The original composite axion model
in Ref.\,\cite{Choi:1985cb}
(see Sec.\,\ref{sec:CAM})
is one of the examples. 

In this class of models, 
extensions have been proposed to 
solve the axion quality problem
by adding a new chiral gauge symmetry to the original composite axion model~\cite{Randall:1992ut, Contino:2021ayn}, by replacing axicolor dynamics with chiral axicolor dynamics~\cite{Gavela:2018paw}, or by incorporating supersymmetry~\cite{Lillard:2018fdt}. 
Note that $\SU(3)_\mathrm{QCD}$ does not mix with the other gauge groups when PQ breaking occurs.
By construction, the PQ symmetry is only anomalous with respect to QCD. 
Therefore, in this class of models, there are no additional instanton effects on the axion potential other than those from QCD. 
This conclusion is different from the case in the CAA model discussed in the previous section, where the constrained instantons have small but non-vanishing contributions to the axion mass.

%-----------------------------------------------------
\subsection{Axial \texorpdfstring{$\boldsymbol{\SU(3)\times [\SU(N)]^{n'_s} \times [\SU(m)]^{n'_s}}$}{} Model}
%-----------------------------------------------------
In Sec.\,\ref{sec:CAAmodel}, we discussed the CAA model, which features $\SU(3)\times [\SU(N)]^{n_s} \times [\SU(4)]^{n_s-1}$ gauge groups. 
In this model, the $\SU(3)$ and $\SU(4)$ groups are embedded in the vector-like subgroups of the $\SU(4)_L \times \SU(4)_R$ flavor symmetry, associated with each axicolor $\SU(N)$ gauge dynamics. 
In contrast, Ref.\,\cite{Redi:2016esr} also proposes another model with $\SU(3)\times [\SU(N)]^{n'_s} \times [\SU(m)]^{n'_s}$ gauge groups, where $\SU(m)$ and $\SU(3)$ are embedded in the axial part of the flavor symmetries of the axicolor $[\SU(N)]^{n_s'}$.

For $n'_s=1$, for example, the model includes left-handed Weyl fermions with the following gauge charges:
\begin{align}
\label{eq:chiral gauging}
(\mathbf{N},\mathbf{m},\mathbf{1})
\oplus
(\mathbf{N},\overline{\mathbf{{m}}},\mathbf{1})
\oplus (\overline{\mathbf{N}},\mathbf{1},\mathbf{3})
\oplus
(\overline{\mathbf{N}},\mathbf{1},\overline{\mathbf{3}})
\oplus
(\overline{\mathbf{N}},\mathbf{1},\mathbf{1})\times 2(m-3)\,,
\end{align}
where the first, second, and third entries in the parentheses denote the representations under the $\SU(N)$, $\SU(m)$, and $\SU(3)$ gauge groups, respectively. 
The fundamental and anti-fundamental fermions of $\SU(N)$ exhibit $\SU(2m)_L$ and $\SU(2m)_R$ flavor symmetries, respectively. 
The $2m$ fundamental fermions are decomposed into $\mathbf{m} \oplus \overline{\mathbf{m}}$ representations of $\SU(m)\subset \SU(2m)_L$, while the $2m$ anti-fundamental fermions are decomposed into $\mathbf{3} \oplus \overline{\mathbf{3}} \oplus (\mathbf{1}\times 2(m-3))$ representations of $\SU(3) \subset \SU(2m)_R$.

This model can be 
derived by deforming the $\SU(3)\times [\SU(N)]^{n_s} \times [\SU(4)]^{n_s-1}$ model in Sec.\,\ref{sec:CAAmodel}, with $n_s = 2 n_s'$.
Specifically, we pair $i$-th $\SU(N)$ with $(n_s-i+1)$-th $\SU(N)$ ($i=1\cdots n'_s$),
and identify one of the $\SU(N)$ symmetries with the complex conjugate of the other.
In Fig.\,\ref{fig:moose axial gauging}, we illustrate this deformation with moose diagrams of the $\SU(3)\times [\SU(N)]^{n_s} \times [\SU(4)]^{n_s-1}$ model and the axial $\SU(3)\times [\SU(N)]^{n'_s} \times [\SU(4)]^{n'_s}$ model, specifically for $n_s = 2$.

In this model, the low energy QCD emerges as the diagonal subgroup of $\SU(m)$'s and $\SU(3)$ as in the case of the model in Sec.\,\ref{sec:CAAmodel}.
The composite axion appears as one of the Goldstone modes, resulting from the spontaneous breaking of the generators of $\SU(2m)_R$ which commute with $\SU(3)_\mathrm{QCD}$.

Since $\SU(3)_\mathrm{QCD}$ appears as the diagonal subgroup of $\SU(m) \times \SU(3)$, constrained instantons can contribute to the axion potential. 
Similar to the CAA model in Sec.\,\ref{sec:CAAmodel}, we find that the model possesses U$(1)$ symmetries that are free from the axicolor anomaly but are anomalous with respect to $\SU(m)$ and $\SU(3)$. 
For example, consider the U$(1)$ symmetry with the charge assignment indicated by the subscripts:
\begin{align}
(\mathbf{N},\mathbf{m},\mathbf{1})_{-1}
\oplus
(\mathbf{N},\overline{\mathbf{m}},\mathbf{1})_{-1}
\oplus (\overline{\mathbf{N}},\mathbf{1},\mathbf{3})_{1}
\oplus
(\overline{\mathbf{N}},\mathbf{1},\overline{\mathbf{3}})_{1}
\oplus
(\overline{\mathbf{N}},\mathbf{1},\mathbf{1})_{1} \times 2(m-3)\ .
\end{align}
This U$(1)$ symmetry remains unbroken and is anomalous with respect to $\SU(m)$ and $\SU(3)$. 
Note that this anomalous but unbroken U$(1)$ symmetry corresponds exactly to U$(1)_1$ as discussed in Tab.\,\ref{tab:model n2} in the previous section.

By revisiting the discussion in Sec.\,\ref{sec:vanishing Effects}, which is based on U$(1)_1$, we find that small constrained instantons affect the axion potential only when the winding numbers of $\SU(m)$'s and $\SU(3)$ are identical. Furthermore, the arguments in Sec.\,\ref{sec:non-vanising Efects} using 't Hooft operators demonstrate that constrained instantons which are significantly smaller than the inverse of the dynamical scale are irrelevant for the axion potential. 
Consequently, we again find that the small instanton effects contribute to the axion potential at most as follows:
\begin{align}
V(a)|_{\mathrm{small\,instanton}} \sim
\Lambda^4 e^{-\frac{8\pi^2}{g^2_\mathrm{QCD}(\Lambda)}}\cos\left(\frac{2Na}{F_a}\right)\ ,
\end{align}
which is subdominant compared with the axion potential generated by low-energy QCD.

\begin{figure}[t]
\centering
\begin{tikzpicture}[scale=1.2]

    \tikzset{t/.style={draw,circle,minimum width=30,opacity=0
    }}
    \tikzset{c/.style={draw,circle,minimum width=25
    }}
    
    \begin{scope}[xscale=1,yscale=1]
     
        \node [c](n1) at (0,.7){$N$};
        \node [c](n2) at (0,-.7){$N$};
        \node [c](4) at (1.5,0){$4$};
        \node [c](3) at (-1.5,0){$3$};
    % transparent nodes, preventing arows from touching nodes   
        \node [t](n1') at (n1){};
        \node [t](n2') at (n2){};
        \node [t](4') at (4){};
        \node [t](3') at (3){};
    % endpoints for up and down arrows
        \node [](up) at (0,1.7){};
        \node [](lo) at (0,-1.7){};
    
        \biarrow (4')-- (n1');
        \biarrow (n1')--(3');
        \biarrow (3')--(n2');
        \biarrow (n2')--(4');
        \antiarrow (n1')--(up);
        \fundarrow (lo)--(n2');
    \end{scope}
    \begin{scope}[xscale=1,yscale=1,shift={(5,0)}]
        \node [c](n) at (0,0){$N$};
        \node [c](4) at (1.5,0){$4$};
        \node [c](3) at (-1.5,0){$3$};    
        \node [t](n') at (n){};
     % transparent and shifted nodes, preventing arrows from overlapping  
        \node [t](nup) at (0,.1){};
        \node [t](4up) at (1.5,.1){};
        \node [t](3up) at (-1.5,.1){};
        \node [t](nlo) at (0,-.1){};
        \node [t](4lo) at (1.5,-.1){};
        \node [t](3lo) at (-1.5,-.1){};
    % endpoints for up and down arrows        
        \node [](up) at (0,1){};
        \node [](lo) at (0,-1){};
        
        \biarrow (4up)--(nup);
        \biarrow (nup)--(3up);
        \antiarrow   (n')--(up);
    % colored (inverted) arrows
        \inwardarrow (nlo)--(3lo);
        \outwardarrow(nlo)--(4lo);
        \antiarrowcolored (n')--(lo);
        
    \end{scope}
    \end{tikzpicture}
    \caption{The moose diagrams for the $\SU(3)\times [\SU(N)]^{n_s} \times [\SU(4)]^{n_s-1}$ model in Sec.\,\ref{sec:CAAmodel} (see Tab.\,\ref{tab:model n2}), and for the axial $\SU(3)\times [\SU(N)]^{n'_s} \times [\SU(4)]^{n'_s}$ model when $n_s = 2$ and $m=4$. An arrow \protect\tikz \protect\draw[->,>={Classical TikZ Rightarrow[scale=2.2]}] (0,0)--(0.5,0); represents the fundamental representation, and an arrow \protect\tikz \protect\draw[>-,>={Classical TikZ Rightarrow[scale=2.2]}] (0,0)--(0.5,0); represents the antifundamental representation. 
    From the $\SU(3)\times [\SU(N)]^2 \times \SU(4)$ model,
    the axial $\SU(3)\times \SU(N) \times \SU(4)$ model is derived by identifying the upper $\SU(N)$ with the complex conjugate of the lower $\SU(N)$, which is indicated by flipping the arrow heads colored in red.  Extending these models to cases with $n'_s > 1$ is straightforward.}
 \label{fig:moose axial gauging}
\end{figure}

%-----------------------------------------------------
\section{Conclusions}
\label{sec:conclusion}
%-----------------------------------------------------
In this paper, we have discussed the small instanton effects in the composite accidental axion (CAA) models
proposed in Ref.\,\cite{Redi:2016esr}, where the U(1)$_\mathrm{PQ}$ symmetry emerges as an accidental global symmetry.
In these models, $\SU(3)_\mathrm{QCD}$ emerges as an unbroken diagonal subgroup of the product gauge group along with the spontaneous breaking of PQ symmetry by axicolor dynamics.
The models feature instanton configurations which do not appear in the low-energy QCD, 
with sizes smaller than the inverse of the PQ symmetry breaking scale $f_a\sim\Lambda$.
In some axion models, these configurations can significantly enhance the axion mass compared with the QCD effects.
Indeed, the axion masses can be significantly enhanced by small instanton effects in models where the axions directly couple to the product gauge groups from which QCD appears as the unbroken subgroup~\cite{Agrawal:2017ksf}, as we have seen in Sec.\,\ref{sec:without fermion SIE}.

In the CAA models, 
there exist small instanton configurations in each $\SU(3)$ and $\SU(4)$ gauge group which do not appear in QCD.
We have confirmed that small instantons in the broken part of the product group do contribute to the axion mass independently of those in QCD.
However, we have also found that these contributions to the axion mass are negligibly small compared to the QCD contributions.
As a result, the axion mass is not enhanced by the small instanton effects in the CAA models.

The crucial feature of the CAA models is that the models possess global chiral U(1) symmetries which are not broken spontaneously but are anomalous with respect to the weakly coupled gauge groups, such as $\SU(3)_\mathrm{w} \times [\SU(4)_\mathrm{w}]^{n_s-1}$.
These U(1) symmetries, which correspond to U(1)$_1$ in Tab.\,\ref{tab:model n2} for the $n_s=2$ model and to U(1)$_1\times$U(1)$_2$ in Tab.\,\ref{tab:model n3} for the $n_s=3$ model, play a crucial role in solving the strong CP problem by eliminating $\theta$-terms in the hidden sectors. 
In generic $n_s$-site models, there are $n_s-1$ such U(1) symmetries.
In generic $n_s$-site models, there are $n_s-1$ such U(1) symmetries.

In this work, we have shown that these U(1) symmetries are also essential in suppressing the small instanton effects.
In the CAA models, small instantons do not contribute to the axion potential unless winding numbers of all the $\SU(3)$ and $\SU(4)$ sectors are identical. 
This occurs because the fermion zero modes around the small instantons cannot be closed due to those chiral U(1) symmetries. 
In the $n_s=2$ model for example, non-vanishing contributions require multiple small instantons in a connected diagram to close up the fermion zero modes, as shown in Fig.\,\ref{fig:twoinstanton in CAA}.
As a result, small instantons with identical winding number for all the unconfined sectors contribute collectively to the axion potential. Thus, the small instantons from the larger coupling sector are accompanied by weakly-coupled small instantons from other sectors.
Consequently, the small instanton effects do not enhance the axion mass, as they are suppressed by a collective instanton factor, $e^{-|m|\times 8\pi^2/g^2_{3\mathrm{w}}}\times e^{-|n_1|\times 8\pi^2/g^2_{4\mathrm{w},1}} \times \cdots\times e^{-|n_{n_s-1}|\times 8\pi^2/g^2_{4\mathrm{w},n_s-1}}$. 
With the same winding numbers from all the sectors, $m=n_1=\cdots=n_{n_s-1}$, this coincides with $e^{-|m|\times 8\pi^2/g^2_{\mathrm{QCD}}}$ due to
the matching condition of the QCD coupling in Eq.\,\eqref{eq:matching condition}.
These small instanton effects are dominated by those with sizes around the inverse of the PQ breaking scale, $\Lambda^{-1}$.
As a result, we find that small constrained instanton effects on the axion mass are at most comparable to the effects of QCD instantons of the same size, $\rho_\mathrm{QCD}\sim\Lambda^{-1}$.
Therefore, they are negligible compared with the low-energy QCD effects.

In conclusion, we emphasize that the presence of anomalous U(1) symmetries, which are not spontaneously broken, significantly restricts the impact of small instanton effects on the axion potential. 
This finding is crucial for understanding the influence of small instantons on the axion potential across a wide range of axion models with high-quality PQ symmetry, not just limited to the CAA models.
It highlights the challenge of increasing the axion mass without spoiling the high-quality PQ symmetry.

%-----------------------------------------------------
\section*{Acknowledgements}
%-----------------------------------------------------
TA acknowledges Sungwoo~Hong and Ryosuke~Sato for valuable communications.
This work is supported by Grant-in-Aid for Scientific Research from the Ministry of Education, Culture, Sports, Science, and Technology (MEXT), Japan, 20H01895 and 20H05860  (S.S.),
 21H04471 and 22K03615 (M.I.) and by World Premier International Research Center Initiative (WPI), MEXT, Japan. 
This work is supported by JST SPRING Grant Number JPMJSP2108 (K.W.).
This work is also supported by FoPM, WINGS Program, the University of Tokyo (T.A.).

\appendix
%-----------------------------------------------------
\section{Notation and Remarks on Euclidean Space}
\label{app:notations}
%-----------------------------------------------------

%-----------------------------------------------------
\subsection{Notation in Euclidean Space}
\label{app:notations, basics in Euclidean space}
%-----------------------------------------------------
We adopt the following notation for coordinates and derivatives in
Minkowski spacetime and Euclidean space related via
\begin{align}
    \xE^{i} = x^i \ , 
    \quad
    \partialE^i = \partial_i\ , \quad
    \xE^4 = i x^0 \ , 
    \quad 
    \partialE^4 = -i \partial_0\ ,
\end{align}
for $i=1,2,3$.
The metric tensors are defined by,
\begin{align}
    g_{\mu\nu} = (+1,-1,-1,-1) \ ,
\end{align}
in Minkowski spacetime and 
\begin{align}
    \gE_{\mu\nu} = (+1,+1,+1,+1) \ ,
\end{align}
in Euclidean space.
The gauge potentials are also related via
\begin{align}
    \AE^i  = A_i\ ,\quad \AE^4 = - i A_0 \ .
\end{align}

In Minkowski spacetime, we use
\begin{align}
(\sigmaM^\mu)_{\alpha \dot{\alpha}}
    = (\mathbb{1}_2, \boldsigma)\ , \quad
    (\barsigmaM^\mu)^{\dot{\alpha} {\alpha}}
    = (\mathbb{1}_2, -\boldsigma)\ , \quad (\mu = 0,1,2,3)\ ,
\end{align}
while we use
\begin{align}
    (\sigmaE^\mu)_{\alpha\dot\alpha}
    =(\boldsigma,i\mathbb{1}_2)\ , \quad  (\barsigmaE^\mu)^{\dot\alpha\alpha}\      =(\boldsigma,-i\mathbb{1}_2)\ , \quad (\mu = 1,2,3,4)
\end{align}
in Euclidean space as in Ref.\,\cite{Espinosa:1989qn}.
Here, $\boldsigma$ denote Pauli matrices.
The generators of the Lorentz group,
$\barsigma_{\mu\nu}$, are defined by 
\begin{align}
    \sigma_{\mu\nu} 
    = \frac{1}{4i} (\barsigma_\mu \sigma_\nu - \barsigma_\nu \sigma_\mu)\ , 
    \quad
     \barsigma_{\mu\nu} 
    = \frac{1}{4i} (\sigma_\mu \barsigma_\nu - \sigma_\nu \barsigma_\mu)\ .
\end{align}
We also define the corresponding generators in Euclidean space similarly from $\sigmaE^\mu$ and $\barsigmaE^\mu$.

For the generators of the $\SU(2)$ gauge group,
we use the Pauli matrices $\boldtau$.
To describe the instanton solution, we use 
\begin{align}
    (\tau^\mu)
    =(\boldtau,\mathbb{1}_2)\ , \quad  (\bartau^\mu)\ 
    =(\boldtau,-i\mathbb{1}_2)\ , \quad (\mu = 1,2,3,4)\ ,
\end{align}
and 
\begin{align}
\label{eq:generators}
    \tau_{\mu\nu} 
    = \frac{1}{4i} (\bartau_\mu \tau_\nu - \bartau_\nu \tau_\mu)\ ,\quad
    \bartau_{\mu\nu} 
    = \frac{1}{4i} (\tau_\mu \bartau_\nu - \tau_\nu \bartau_\mu)\ .
\end{align}

Finally, 
we use the following notation for 
the two-dimensional anti-symmetric tensors,
\begin{align}
    \epsilon^{12}&=-
    \epsilon^{21}=1\ , \quad
    \epsilon_{12}  = - \epsilon_{21} = -1 \ , \quad
    \epsilon_{1}{}^{2} = - 
    \epsilon_{2}{}^{1} = 1 \ .
\end{align}

%-----------------------------------------------------
\subsection{Remarks on Fermions in Euclidean Space}
\label{app:remarks on fermions in Euclidean space}
%-----------------------------------------------------
We make some remarks on the relationship between fermions in Minkowski spacetime and Euclidean space.
As an example, we consider the Lagrangian in Minkowski spacetime,
\begin{align}
\label{eq:Minkowski Lagrangian}
\mathcal{L}_\mathrm{M}
    =  (\chi_L)^\dagger_{\dot{\alpha}} i(\barsigmaM^\mu)^{\dot{\alpha}\alpha} \ \partial_\mu (\chi_L)_\alpha 
+ (\bar{\eta}_R)^{\alpha} i(\sigmaM^\mu)_{\alpha \dot{\alpha}} \ \partial_\mu (\bar{\eta}_R)^{\dagger\dot{\alpha}}    
- m (\chi_L)_\alpha (\bar{\eta}_R)^\alpha 
    - m (\chi_L)^{\dagger}_{\dot\alpha}
    (\bar{\eta}_R)^{\dagger\dot\alpha}\, .
\end{align}
Here, both $(\chi_L)_\alpha$ and $(\bar\eta_R)^{\alpha}$ are left-handed Weyl spinors.
We follow the conventions of the spinor indices in
Minkowski spacetime in Ref.\,\cite{Dreiner:2008tw}.

In Minkowski spacetime, ${\chi}^\dagger_L$ 
and $\bar\eta^\dagger_R$ are the hermitian conjugates 
of ${\chi}_L$ and $\bar\eta_R$.
In Euclidean space, on the other hand, 
(${\chi}^\dagger_L$, $\bar\eta^\dagger_R$)
and (${\chi}_L$, $\bar\eta_R$) should be translated into independent fermions, respectively.
We 
introduce the subscripts $A$ and $B$ to distinguish the dotted and undotted spinors in Euclidean space,
which are related to the left-handed
and right-handed Weyl fermions (i.e., undotted and the dotted fermions) 
in Minkowski spacetime via,
\begin{align}
\label{eq:fermions in M/E}
    (\chi_L)^{\dagger}_{\dot{\alpha}} \rightarrow (\chi_A^{\dagger})_{\dot{\alpha}}\ , \,\,\, (\chi_L)_\alpha \rightarrow (\chi_B)_\alpha\ , \,\,\,
    (\bar\eta_R)^{\dagger\dot{\alpha}} \rightarrow (\bar\eta_A^{\dagger})^{\dot{\alpha}}\ , \,\,\,
    (\bar\eta_R)^{\alpha} \rightarrow (\bar\eta_B)^\alpha\ .
\end{align}
This notation is following Ref.\,\cite{Espinosa:1989qn}.
In the main text, $L/R$, $A/B$ and $\mathrm{M}/\mathrm{E}$ subscripts are omitted since they should not cause any particular confusion.

Let us comment on this notation. 
In Euclidean space, the space rotation is $\SO(4) \simeq \SU(2)_A \times \SU(2)_B$. 
The subscripts $A$ or $B$ of the fermions indicate which subgroup of spatial rotation is associated with the fermion, $\SU(2)_A$ or $\SU(2)_B$. 
Note again that fermions labeled with subscripts $A$ and $B$ are not related by Hermitian conjugation, and the $\dagger$ symbol should be considered as part of the fermion's name rather than denoting a conjugate.

In this notation, the Minkowski Lagrangian \eqref{eq:Minkowski Lagrangian} corresponds to the Euclidean Lagrangian,
\begin{align}
 \LE = -(\chi_A^\dagger)_{\dot{\alpha}} i(\barsigmaE^\mu)^{\dot{\alpha}\alpha} \ \partial_\mu (\chi_B)_\alpha\ 
+(\bar\eta_B)^{\alpha} i(\sigmaE^\mu)_{\alpha \dot{\alpha}} \ \partial_\mu (\bar\eta_A^\dagger)^{\dot{\alpha}} 
    - m (\chi_B)^\alpha (\bar\eta_B)_{\alpha} 
    - m (\chi_A^{\dagger})_{\dot\alpha}
    (\bar\eta_A^\dagger)^{\dot{\alpha}}\ .
\end{align}
Here, raising and lowering indices $\alpha$ and $\dot{\alpha}$ are defined by multiplying the antisymmetric epsilon symbol as in Minkowski spacetime.
An extra minus sign in front of $\barsigmaE$ 
is relevant to reproduce the Minkowski propagators by Wick rotation.

We use the following notation,
\begin{align}
    \bar{\slashed{\partial}}_E = - \barsigmaE^\mu\partial_\mu\ , \quad
    \slashed{\partial}_E= \sigmaE^\mu \partial_\mu\ ,
\end{align}
and also $\bar{\slashed{D}}_\mathrm{E}$ and $\slashed{D}_\mathrm{E}$ for the covariant derivatives in gauge theories.

%-----------------------------------------------------
\section{Zero Modes of Massive Fermions around Constrained Instanton}
\label{app:zero mode}
%-----------------------------------------------------
We briefly summarize the features of fermion zero modes around the constrained instantons.
As in Sec.\,\ref{sec:constrained instanton}, let us consider an $\SU(2)$ gauge theory in Euclidean space. 
We introduce two pairs of $\SU(2)$ singlet fermions $(\bar{e}_A^\dagger,\bar{e}_B)$ and $(\bar\nu_A^\dagger,\bar\nu_B)$,
and a pair of  $\SU(2)$ doublet fermions $(\ell_A^\dagger,\ell_B)$.
Note that each pair $(f_A^\dagger,f_B)$
is mapped to $f_L^\dagger$ and its 
conjugate $f_L$ in Minkowski spacetime
(see Eq.\,\eqref{eq:fermions in M/E}). 
Here, following the notation in Ref.\,\cite{Espinosa:1989qn}, the fermions are named like the SM leptons
(and the right-handed neutrino), although they are not related to the SM leptons.

We consider the following Lagrangian,
\begin{equation}
\begin{split}
    \LE &= \dfrac{1}{2 g^2}\mathrm{Tr}(F_{\mu\nu}F_{\mu\nu}) + (D_{\mu}H)^{\dagger}(D_{\mu}H) + \dfrac{\lambda}{4}(H^{\dagger}H - v^2)^2
    \\&- \ell_A^\dagger i\barsigma_\mu D_\mu \ell_B 
    + \bar{e}_B i\sigma_\mu \partial_\mu \bar{e}_A^\dagger
    + \bar\nu_B i\sigma_\mu \partial_\mu \bar\nu_A^\dagger \\
    &-y_e \ell_A^\dagger H \bar{e}_A^{\dagger} - y_e \bar{e}_B H^\dagger \ell_B  
    -y_\nu \ell_A^\dagger (\epsilon H)^\dagger \bar\nu_A^\dagger 
    - y_\nu \bar\nu_B (\epsilon H) \ell_B\ ,
    \end{split}    
\end{equation}
where $y_{e,\nu}$ denote the Yukawa coupling constants,
and $\epsilon$ denotes the two-dimensional anti-symmetric invariant tensor with respect to $\SU(2)$ gauge group.
At the vacuum, the scalar obtains a VEV, $H=\qty(0,\,v)^T$, and all
the fermions obtain masses, $m_e = y_e v$ and $m_\nu = y_\nu v$, respectively.
The fermion kinetic terms can be rewritten as
\begin{equation}
    \LE^\mathrm{f} = \Psi^\dagger i \hat{\slashed{D}}_H\Psi
    \ ,\
    \Psi^\dagger = 
    \qty(
        \ell_A^\dagger, \bar{e}_B, \bar\nu_B
    )
    \ ,\
    \Psi=
    \begin{pmatrix}
        \ell_B\\
        \bar{e}_A^{\dagger} \\
        \bar{\nu}_A^{\dagger}
    \end{pmatrix}
    \ ,  
\end{equation}
where we have defined the derivative operator,
\begin{align}
i \hat{\slashed{D}}_H \equiv 
    \begin{pmatrix}
        -i \barsigma_\mu D_\mu^\mathrm{inst} & -y_e H^\mathrm{inst} & -y_\nu (\epsilon H^\mathrm{inst})^\dagger \\
        -y_e H^{\mathrm{inst}\dagger} & i \sigma_\mu \partial_\mu & 0 \\
        -y_\nu(\epsilon H^\mathrm{inst}) & 0 &  i \sigma_\mu \partial_\mu 
    \end{pmatrix}\ .
\end{align}
Here, the superscript ``inst" 
denotes the constrained instanton background (see Eqs.\,\eqref{eq:constrained instanton profile of A} and \eqref{eq:constrained instanton profile of H}).
Note that 
the off-diagonal elements give the  mass terms for the fermions at far from the instanton, i.e., 
$H^\mathrm{inst}|_{x\to \infty}\to (0,v)^T$.

As discussed in Ref.\,\cite{Espinosa:1989qn}, $i\hat{\slashed{D}}_H$ has a normalizable zero mode in the anti-instanton background which behaves as
\begin{align}
    {\ell_B{}_{\alpha i}(x)} &= x_\mu (\sigma_\mu)_{\alpha\dot\alpha}\left[\mathcal{N} \dfrac{\rho}{x(x^2+\rho^2)^{3/2}} \epsilon^{\dot\alpha}{}_{i}
    + \order{(\rho m_{A,H,e,\nu})^2}\right]\ ,
    \\
    {\bar{e}_A^{\dagger}{}^{\dot\alpha}(x)} &=
    -\dfrac{i}{2}\mathcal{N} \rho m_e \dfrac{1}{x^2+\rho^2} \delta^{\dot\alpha}{}_{1}
    + \order{(\rho m_{A,H,e})^2}\ ,
    \\
    {\bar{\nu}_A^{\dagger}{}^{\dot\alpha}(x)} &=
    +\dfrac{i}{2}\mathcal{N}\rho m_\nu \dfrac{1}{x^2+\rho^2} \delta^{\dot\alpha}{}_{2}
    + \order{(\rho m_{A,H,\nu})^2}\ ,
\end{align}
at $x\ll m_{A,H,e,\nu}^{-1}$.
They decay
exponentially at $x \gg  m_{A,H,e,\nu}^{-1}$.
Here, $i$ denotes the gauge $\SU(2)$ index of the doublet fermion, and $\mathcal{N}$ is a finite normalization constant.
In the instanton background, on the other hand, 
there are zero modes in 
$\ell_A^\dagger$ with $\order{\rho m_{e,\nu}}$ 
contributions of $e_{B}^{\dagger}$ and $\nu_{B}^{\dagger}$.
In Minkowski spacetime, the zero mode around the constrained anti-instanton appears in $\ell_L$ (with $\bar{e}_R^\dagger/\bar{\nu}_R^\dagger$)
and the zero mode around the constrained instanton in 
 $\ell_L^\dagger$ (with $\bar{e}_R/\bar{\nu}_R$).

\bibliographystyle{apsrev4-1}
\bibliography{bibtex}

\end{document}